\documentclass{aa}  

\usepackage{color}
\usepackage{graphicx}
\usepackage[varg]{txfonts}
\usepackage{natbib}
\bibpunct{(}{)}{;}{a}{}{,}
\begin{document}
\newcommand{\Euppk}{\mbox{$E_{\rm upp}/k$}}
\newcommand{\LSR}{\mbox{\em LSR}}
\newcommand{\vycma}{\object{VY\,CMa}}
\newcommand{\tmb}{\mbox{$T_{\rm mb}$}}
\newcommand{\jk}{\mbox{$J_K$}}
\newcommand{\jn}{\mbox{$J_N$}}
\newcommand{\jkl}[2]{\mbox{$#1_{#2}$}}
\newcommand{\nuuu}{\mbox{$v_1$=1}}
\newcommand{\nudu}{\mbox{$v_2$=1}}
\newcommand{\nudd}{\mbox{$v_2$=2}}
\newcommand{\nutu}{\mbox{$v_3$=1}}
\newcommand{\nuz}{\mbox{$v$=0}}
\newcommand{\nhtzz}{\mbox{$0^-_0$}}
\newcommand{\nhtuz}{\mbox{$1^+_0$}}
\newcommand{\nhttz}{\mbox{$3^+_0$}}
\newcommand{\nhtdz}{\mbox{$2^-_0$}}
\newcommand{\nhttu}{\mbox{$3^-_1$}}
\newcommand{\nhtdu}{\mbox{$2^+_1$}}
\newcommand{\nhttd}{\mbox{$3^+_2$}}
\newcommand{\nhtdd}{\mbox{$2^-_2$}}
\newcommand{\jkk}{\mbox{$J_{K_{\rm a},K_{\rm c}}$}}
\newcommand{\jkul}[4]{\mbox{${#1}_{#2}$--${#3}_{#4}$}}
\newcommand{\jkkul}[6]{\mbox{$#1_{#2,#3}$--$#4_{#5,#6}$}}
\newcommand{\jkkl}[3]{\mbox{$#1_{#2,#3}$}}
\newcommand{\twopihalf}{\mbox{$^{2}\Pi_{1/2}$}}
\newcommand{\twopithalf}{\mbox{$^{2}\Pi_{3/2}$}}
\newcommand{\nutwo}{\mbox{$\nu_2$}}
\newcommand{\jtmdm}{\mbox{$J$=$3/2$--$1/2$}}
\newcommand{\rottenegg}{\mbox{OH\,231.8+4.2}}
\newcommand{\qxpup}{\mbox{QX\,Pup}}
\newcommand{\doceCO}{\mbox{$^{12}$CO}}
\newcommand{\sotwo}{\mbox{SO$_2$}}
\newcommand{\treceCO}{\mbox{$^{13}$CO}}
\newcommand{\CdieciochoO}{\mbox{C$^{18}$O}}
\newcommand{\HtreceCN}{\mbox{H$^{13}$CN}}
\newcommand{\treceCN}{\mbox{$^{13}$CN}}
\newcommand{\CdoO}{C\mbox{$^{18}$O}}
\newcommand{\jdn}{\mbox{$J$=10$-$9}}
\newcommand{\jdsq}{\mbox{$J$=16$-$15}}
\newcommand{\jqc}{\mbox{$J$=15$-$14}}
\newcommand{\jsc}{\mbox{$J$=6$-$5}}
\newcommand{\jno}{\mbox{$J$=9$-$8}}
\newcommand{\jtd}{\mbox{$J$=3$-$2}}
\newcommand{\jcc}{\mbox{$J$=5$-$4}}
\newcommand{\jdu}{\mbox{$J$=2$-$1}}
\newcommand{\juc}{\mbox{$J$=1$-$0}}
\newcommand{\ammonia}{\mbox{NH$_3$}}
\newcommand{\vosio}{\mbox{$^{28}$SiO}}
\newcommand{\vnsio}{\mbox{$^{29}$SiO}}
\newcommand{\trsio}{\mbox{$^{30}$SiO}}
\newcommand{\sodos}{\mbox{SO$_2$}}
\newcommand{\tcsodos}{\mbox{$^{34}$SO$_2$}}
\newcommand{\protonwater}{\mbox{H$_3$O$^+$}}
\newcommand{\water}{\mbox{H$_2$O}}
\newcommand{\pwater}{\mbox{para-H$_2$O}}
\newcommand{\owater}{\mbox{ortho-H$_2$O}}
\newcommand{\waterdo}{\mbox{H$_2^{18}$O}}
\newcommand{\pwaterdo}{\mbox{para-H$_2^{18}$O}}
\newcommand{\owaterdo}{\mbox{ortho-H$_2^{18}$O}}
\newcommand{\waterds}{\mbox{H$_2^{17}$O}}
\newcommand{\owaterds}{\mbox{ortho-H$_2^{17}$O}}
\newcommand{\pwaterds}{\mbox{para-H$_2^{17}$O}}
\newcommand{\hydroxyl}{\mbox{OH}}
\newcommand{\rotten}{\mbox{H$_2$S}}
\newcommand{\gsim}{\raisebox{-.4ex}{$\stackrel{>}{\scriptstyle \sim}$}}
\newcommand{\lsim}{\raisebox{-.4ex}{$\stackrel{<}{\scriptstyle \sim}$}}
\newcommand{\psim}{\raisebox{-.4ex}{$\stackrel{\propto}{\scriptstyle \sim}$}}
\newcommand{\kms}{\mbox{km~s$^{-1}$}}
\newcommand{\s}{\mbox{$''$}}
\newcommand{\mloss}{\mbox{$\dot{M}$}}
\newcommand{\msun}{\mbox{$M_{\mbox \sun}$}}
\newcommand{\lsun}{\mbox{$L_{\mbox \sun}$}}
\newcommand{\my}{\mbox{$M_{\odot}$~yr$^{-1}$}}
\newcommand{\ls}{\mbox{$L_{\odot}$}}
\newcommand{\ms}{\mbox{$M_{\odot}$}}
\newcommand{\mm}{\mbox{$\mu$m}}
\def\arcdeg{\hbox{$^\circ$}}
\newcommand{\seca}{\mbox{\rlap{.}$''$}}
\newcommand{\dega}{\mbox{\rlap{.}$^\circ$}}
\newcommand{\aprop}{\raisebox{-.4ex}{$\stackrel{\propto}{\scriptstyle\sf \sim}$}}
\newcommand{\apropg}{\raisebox{-.4ex}{$\stackrel{\Large \propto}{\sim}$}}
\newcommand{\about}{\mbox{$\sim$}}
\newcommand{\ttt}[1]{\mbox{10$^{#1}$}}
\newcommand{\hspec}{7.715cm}
\newcommand{\hdoublespec}{16.99cm}

   \title{HIFISTARS
   Herschel/HIFI\thanks{Herschel is an ESA space observatory with science
   instruments provided by European-led Principal Investigator consortia and 
   with important participation from NASA. HIFI is the Herschel Heterodyne 
   Instrument for the Far Infrared.}
   observations of VY\,Canis Majoris\thanks{Appendices A and B
   are only available in electronic form at http://www.edpsciences.org}}

   \subtitle{Molecular-line inventory of the envelope around the largest known star}

   \author{J.\,Alcolea\inst{1}
          \and 
          V.\,Bujarrabal\inst{2}
          \and 
          P.\,Planesas\inst{1} 
          \and 
          D.\,Teyssier\inst{3}
          \and 
          J.\,Cernicharo\inst{4} 
          \and
          E.\,De\,Beck\inst{5,6} 
          \and
          L.\,Decin\inst{5,7} 
          \and 
          C.\,Dominik\inst{7,8}
          \and  
          K.\,Justtanont\inst{9} 
          \and
          A.\,de Koter\inst{5,7}
          \and 
          A.P.\,Marston\inst{3}
          \and 
          G.\,Melnick\inst{10}
          \and 
          K.M.\,Menten\inst{6} 
          \and 
          D.A.\,Neufeld\inst{11} 
          \and 
          H.\,Olofsson\inst{9} 
          \and 
          M.\,Schmidt\inst{12} 
          \and 
          F.L.\,Sch\"oier\inst{9}$^{,}$\thanks{Deceased 14 January 2011} 
          \and 
          R.\,Szczerba\inst{12} 
          \and 
          L.B.F.M.\,Waters\inst{7,13}             
          }
   	\institute{
   	Observatorio Astron\'omico Nacional (IGN), Alfonso XII, 3 y 5, E-28014 Madrid (Spain); 
    \email{j.alcolea[at]oan.es}
	\and
	Observatorio Astron\'omico Nacional (IGN), PO Box 112, E-28803 Alcal\'a de Henares (Spain) 
	\and
	European Space Astronomy Centre, Urb. Villafranca del Castillo, PO Box 50727, E-28080 Madrid (Spain) 
	\and
	CAB, INTA-CSIC, Ctra. de Torrej\'on a Ajalvir, km 4, E-28850 Torrej\'on de Ardoz (Spain) 
	\and
	Instituut voor Sterrenkunde, Katholieke Universiteit Leuven, Celestijnenlaan 200D BUS 2401, 3001 Leuven (Belgium) 
	\and
	Max-Planck-Institut f\"ur Radioastronomie, Auf dem H{\"u}gel 69, 53121 Bonn (Germany) 
	\and
	Sterrenkundig Instituut Anton Pannekoek, University of Amsterdam, Science Park 904, 1098 Amsterdam (The Netherlands) 
	\and
	Department of Astrophysics/IMAPP, Radboud University Nijmegen, Nijmegen  (The Netherlands) 
	\and
    Onsala Space Observatory, Dept. of Earth and Space Sciences, Chalmers University of Technology, 43992 Onsala (Sweden)	
	\and
    Harvard-Smithsonian Center for Astrophysics, Cambridge, MA 02138 (USA) 
	\and
    The Johns Hopkins University, 3400 North Charles St, Baltimore, MD 21218 (USA) 
	\and
    N. Copernicus Astronomical Center, Rabia\'nska 8, 87-100 Toru\'n (Poland) 
	\and
    Netherlands Institute for Space Research, Sorbonnelaan 2, 3584 CA Utrecht (The Netherlands) 
             }
             
\date{Received April 11, 2013; accepted September 25, 2013}

 
  \abstract
   {}
   {The study of the molecular gas in the circumstellar envelopes of evolved stars is normally undertaken by
   observing lines of CO (and other species) in the millimetre-wave
   domain. In general, the excitation requirements of the
   observed lines are low at these wavelengths, and therefore these observations predominantly probe the cold outer envelope while
   studying the warm inner regions of the envelopes normally requires sub-millimetre (sub-mm) and far-infrared (FIR) observational data.}
   {To gain insight into the physical conditions and kinematics of the warm
   (100--1000\,K) gas around the red hyper-giant \vycma, we performed 
   sensitive high spectral resolution observations of molecular lines in the sub-mm/FIR using 
   the HIFI instrument of the 
   Herschel Space Observatory. We observed CO, \water, and other molecular species,
   sampling excitation energies from a few tens to a few thousand K. These observations are part of 
   the Herschel Guaranteed Time Key Program HIFISTARS.}
   {We detected the \jsc, \jdn, and \jdsq\ lines of \doceCO\ and \treceCO\ at $\sim$ 100, 300, and
   750\,K above the ground state (and the \treceCO\ $J$=9--8 line). These lines are crucial for improving the modelling of the internal layers 
   of the envelope around
   \vycma. We also detected 27 lines of \water\ and its isotopomers, and 96 lines of species such as 
   \ammonia, SiO, SO, \sodos\, HCN, OH and others, some of them originating from 
   vibrationally excited levels. Three lines were not unambiguously assigned.} 
   {Our observations confirm that \vycma's envelope must consist
   of two or more detached components. The molecular excitation in the outer layers is significantly lower than
   in the inner ones, resulting in strong self-absorbed profiles in molecular lines that are optically thick
   in this outer envelope, for instance, low-lying lines of \water. Except for the most abundant species, CO and
   \water, most of the molecular emission detected at these sub-mm/FIR wavelengths arise from the central parts
   of the envelope.
   The spectrum of \vycma\ is very prominent in vibrationally excited lines, which are caused by the strong IR pumping 
   present in the central regions. Compared with envelopes of other massive evolved stars, \vycma 's emission is particularly
   strong in these vibrationally excited lines, as well as in the emission from less abundant species such as
   \HtreceCN, SO, and \ammonia.}


   \maketitle
%

\section{Introduction}\label{sec:introduction}

The very luminous red evolved star \vycma\ 
\citep[M2.5 -- M5e Ia;][]{wallerstein1958,humphreys1974}, 
also known as IRC\,$-$30087 and CRL\,1111, is one of the most extreme stars in our galaxy. 
Thanks to trigonometric parallax measurements of \water\ and \vosio\ masers, the distance
to \vycma\ is well known, 1.10--1.25\,kpc \citep[see][]{choi2008,zhang2012}, resulting  
in a total luminosity of 3\,10$^5$\,\lsun\ \citep{smith2001}. This value is close to the empirical 
limit for cool evolved stars \citep{humphreys-davidson1994}, and therefore \vycma\ has been 
classified as a red hyper-giant (RHG), see \citet{jager1998}. These accurate distance estimates, 
in combination with VLTI/AMBER measurements of the angular size of \vycma, 
result in a value for the 
stellar radius that is the highest among well-characterized stars in our galaxy:
1420$\pm$120\,$R_{\sun}$ 
\citep{wittkowski2012}.
Values for its initial mass are more 
uncertain, ranging from 15 to 50\,\msun\ 
\citep[see][]{wittkowski1998,knapp1993,smith2001,choi2008,wittkowski2012}. 
\vycma\ is also characterized by displaying a huge mass loss, with rate values in the 
range of 5\,10$^{-5}$ to 10$^{-4}$\,\my\ \citep{decin2006}, and as high as 2\,10$^{-3}$ 
at least during the past 300\,yr \citep[see][and references therein]{smith2009}; this last value 
is so high that it is thought that \vycma\ might explode as a supernova
from its present state in about 10$^5$ yr \citep[see][]{smith2009}.

Owing to its high mass-loss rate, the star is surrounded by a thick circumstellar envelope,
which was detected as an optical reflection nebulosity almost a century ago 
\citep{perrine1923}, and which has been studied in great detail using the Hubble Space Telescope and 
ground-based adaptive optics \citep[see][and references therein]{smith2001,humphreys2007}. This 
nebulosity is relatively compact, $8\s\!\times8\s$ at most (even in the mid-IR), but 
displays much sub-structure.
At short wavelengths, the nebula is highly asymmetric, with a much brighter lobe located 
south of the assumed position of the star. This suggests the existence of a bipolar structure 
oriented in the north-south direction, the south lobe pointing towards us 
\citep[see][for example]{herbig1970}. Due to the large amount of dust in the envelope, 
the star itself is hardly visible in the optical, and most of its luminosity is re-radiated 
in the mid-IR and at longer wavelengths. This and the high luminosity of the source 
makes \vycma\ one of the brightest IR sources in the sky.

The envelope around \vycma\ has a very rich molecular variety. In the centimetre- and millimetre-wave
range, \vycma\ has been known for many years to exhibit strong maser emission in the three 
classical circumstellar maser species: \hydroxyl, \water, and \vosio. In particular, \vycma\ 
is the strongest emitter of highly excited \vosio\ ($v\!>\!2$) and \water\ ($v_2\!\geq\!1$) masers
\citep[see e.g.][]{pepe1993,menten2006}. At these wavelengths, however, the strength of the 
thermal molecular emission of \vycma's envelope is not outstanding, probably due to its large 
distance and small extent, which led to little interest in the object.
This situation has changed drastically in the past half decade, when the results of 
sensitive spectral surveys in millimetre, sub-millimetre (sub-mm), and far-IR (FIR) ranges were published, from which a 
wealth of molecular lines from about two dozen species have been identified. 
Since then, this source has become the main target of the search of new 
molecular species in O-rich circumstellar environments.

Using the ISO Short Wavelength Spectrometer, \citet{neufeld1999} discovered that the FIR spectrum of 
\vycma\ is rich in spectral features that are caused by water vapour. 
Subsequently, \citet{polehampton2010} studied \vycma\ at longer wavelengths with
the ISO Long Wavelength Spectrometer, finding that it is also dominated by strong lines of \water, and 
concluding that this is the most abundant molecular species after H$_2$ in its circumstellar envelope. In addition
to \water, CO, and \hydroxyl, these authors also reported the identification of lines
from less common species such as \ammonia, and tentatively \protonwater. In the same year, 
\citet{royer2010} published the Herschel PACS and SPIRE full spectral scan (55 to 672\,$\mu$m 
wavelength range). These more sensitive observations confirmed that the far-IR and 
sub-mm spectrum of \vycma\ is dominated by \water\ lines, which are responsible for 
nearly half of the 930 emission lines identified in this work. In addition to \water, lines of 
CO, CN, CS, SO, \ammonia, \hydroxyl, SiO, SiS, \sodos, \rotten, and \protonwater\
(and of some of their isotopomers) are also detected in the spectrum, in spite of
the relatively poor spectral resolution. From a preliminary analysis of the \water\ spectrum, Royer and collaborators 
concluded that the fractional abundance of this species (w.r.t. the total abundance of 
hydrogen) is very high, 3$\times$10$^{-4}$. They also found 
a low ortho-to-para abundance ratio of $\sim$1.3:1, 
which would support the hypothesis that non-equilibrium chemical processes control the formation of \water\ 
in this envelope. 
Meanwhile, \citet[][but see also \citealp{ziurys2007} and \citeyear{ziurys2009}; \citealp{tenenbaum2009} and 
\citeyear{tenenbaum2010a}; \citealt{tenenbaum2010b}]{tenenbaum2010c} 
presented the results of a full-scan survey in the 215--285\,GHz range, conducted with the 
Arizona Radio Obseratory's 10-meter diameter Submillimeter Telescope\footnote{Formerly 
known as the Heinrich Hertz telescope.} on Mt.\,Graham (ARO-10m SMT). 
This survey yielded the detection of ten more new species 
in the envelope of \vycma, namely HCN, HNC, HCO$^{+}$, CN, NS, PN, PO, NaCl, AlO, and AlOH. 
These studies will soon be 
complemented with the results from the 
full spectral surveys that have been or are being performed with Herschel/HIFI and the IRAM-30m 
telescope in all the available bands of these two instruments, 
and with the Submillimeter Array (SMA) in the 280--355\,GHz ($\sim$\,0.9\,mm) frequency range 
\citep[see][]{kaminski2013}. 
Altogether, these 
works have revealed, and will continue to reveal, the chemical richness and complexity that 
can be present in the circumstellar envelopes of cool, high mass-loss O-rich stars.

A proper understanding of the chemical characteristics of the circumstellar
envelope of this unique source necessarily requires a prior good knowledge 
of the main physical conditions in the envelope. For the cool layers, this
can be attained from the ground by means of low-$J$ rotational lines of CO and other
abundant species. However, to gain insight into the deep warm layers where
most of the molecular species are formed, we need to observe high-$J$ lines that
in general are not or not easily accessible for ground-based telescopes. Moreover, for O-rich
environments, the expected high \water\ abundances make this species 
the major coolant for the gas phase, and therefore knowing the distribution and excitation 
of \water\ is crucial for determining the temperature of the molecular gas in
general. Yet, it is important to observe these molecular lines with very high spectral
resolution to adequately probe the velocity field in the envelope.
All these observational needs are fully met by the HIFI instrument on-board
the Herschel Space Observatory. 
Here we present new Herschel/HIFI observations 
of \vycma. Observational and data reduction procedures are detailed in Sect.\,2. 
In Sect.\,3 and the appendices, we 
discuss the main observational results.
The conclusions are presented in Sect.\,4.

\begin{table*}
\caption{Telescope characteristics and main observational parameters.}             
\label{tabteles}              
\begin{tabular}{@{}lccccccccccc@{}}     
\hline\hline       
\multicolumn{1}{c}{Herschel\,\,\,\,}   
                        & Obs. date   & Durat. & \multicolumn{2}{c}{{\small HIFISTARS} setting} & \multicolumn{2}{c}{Sky frequency coverage} 
                                                                         & $T^{\rm DSB}_{\rm sys}$ 
                                                                                         & HIFI 
                                                                         & \multicolumn{3}{c@{}}{{\small HPBW}~ $T_{\rm mb}$/$T^*_{\rm a}$ 
                                                                                                                      ~Cal.$^\S$}\\
\multicolumn{1}{c}{OBSID\,\,\,\,}  
                        & yr:mo:day   & (secs.)  & \multicolumn{2}{c}{name  ~~ LO (GHz)}     
                                                                         &    LSB (GHz)  & USB (GHz) 
                                                                         &      (K)      & band & \multicolumn{1}{c}{~(\arcsec)}  &           & uncer.\\ \hline
1342192528              & 2010:03:21  &   1575   & ~~14     &  ~~564.56    &  ~~556.55--~~560.69 & ~~568.54--~~572.68 
                                                                         &    ~~~~93     & 1B   & ~~37.5 &   1.3  &  15\%\\
1342192529              & 2010:03:21  &   1575   & ~~13     &  ~~614.86    &  ~~606.85--~~610.99 & ~~618.84--~~622.98 
                                                                         &    ~~~~91     & 1B   & ~~34.5 &   1.3  &  15\%\\
1342192533              & 2010:03:21  &   1623   & ~~12     &  ~~653.55    & ~~645.55--~~649.69 & ~~657.54--~~661.68 
                                                                         &     ~~131     & 2A   & ~~32.0 &   1.3  &  15\%\\
1342192534              & 2010:03:21  &  ~~617   & ~~17     &  ~~686.42    & ~~678.42--~~682.56 & ~~690.41--~~694.55 
                                                                         &     ~~142     & 2A   & ~~30.8 &   1.3  &  15\%\\
1342194680${^\dagger}$  & 2010:04:13  &   1618   & ~~11     &  ~~758.89    & ~~750.90--~~755.04 & ~~762.89--~~767.03 
                                                                         &     ~~196     & 2B   & ~~27.8 &   1.3  &  15\%\\
1342195039${^\dagger}$  & 2010:04:18  &   1505   & ~~10     &  ~~975.23    & ~~967.27--~~971.41 & ~~979.26--~~983.40 
                                                                         &     ~~352     & 4A   & ~~21.7 &   1.3  &  20\%\\
1342195040${^\dagger}$  & 2010:04:18  &   1505   & ~~09     &  ~~995.63    & ~~987.67--~~991.81 & ~~999.66--1003.80 
                                                                         &     ~~364     & 4A   & ~~21.3 &   1.3  &  20\%\\
1342194786              & 2010:04:17  &   1487   & ~~08     &   1102.92    &   1094.98--1099.11 &  1106.97--1111.10 
                                                                         &     ~~403     & 4B   & ~~19.2 &   1.3  &  20\%\\
1342194787              & 2010:04:17  &   1487   & ~~07     &   1106.90    &   1098.95--1103.09 &  1110.94--1115.08 
                                                                         &     ~~416     & 4B   & ~~19.1 &   1.3  &  20\%\\
1342192610              & 2010:03:22  &   1618   & ~~06     &   1157.67    &   1149.72--1153.85 &  1161.71--1165.84  
                                                                         &     ~~900     & 5A   & ~~18.3 &   1.5  &  20\%\\
1342192611              & 2010:03:22  &   1538   & ~~05     &   1200.90    &   1192.95--1197.08 &  1204.94--1209.07 
                                                                         &      1015     & 5A   & ~~17.6 &   1.5  &  20\%\\
1342195105${^\ddagger}$ & 2010:04:19  &   3547   & ~~04     &   1713.85    &   1709.11--1711.68 &  1716.37--1718.97 
                                                                         &      1238     & 7A   & ~~12.4 &   1.4  &  30\%\\
1342195106${^\ddagger}$ & 2010:04:19  &   3851   & ~~03     &   1757.68    &   1752.95--1755.52 &  1760.20--1762.77 
                                                                         &      1580     & 7A   & ~~12.0 &   1.4  &  30\%\\
1342230403${^\ddagger}$ & 2011:10:09  &   3103   & ~~\,19$^\flat$     
                                                          &   1766.89    &   1762.23--1764.78 &  1769.48--1772.04 
                                                                         &      1248     & 7A   & ~~12.0 &   1.4  &  30\%\\
1342194782${^\ddagger}$ & 2010:04:17  &   2992   & ~~\,16$^\sharp$     
                                                          &   1838.31    &   1833.59--1836.16 &  1840.84--1843.41 
                                                                         &      1300     & 7B   & ~~11.6 &   1.4  &  30\%\\
1342196570${^\ddagger}$ & 2010:05:15  &   3207   & ~~\,16$^\sharp$     
                                                          &   1838.47    &   1833.73--1836.30 &  1840.99--1843.55 
                                                                         &      1415     & 7B   & ~~11.6 &   1.4  &  30\%\\
1342194781${^\ddagger}$ & 2010:04:17  &   1535   & ~~01     &   1864.82    &   1860.10--1862.67 &  1867.36--1869.92 
                                                                         &      1330     & 7B   & ~~11.4 &   1.4  &  30\%\\
\hline
\end{tabular}
\tablefoot{
$^\S$ Following \citet{teyssier2012}, we based our absolute-calibration accuracy estimates on the error budget reported by 
\citet{roelfsema2012}, plus an additional contribution arising from the ripples in the baseline for bands 7A and 7B. 
${^\dagger}$ Instabilities in the baseline that simultaneously affect both receivers were removed by discarding several sub-scans. 
${^\ddagger}$ Ripples in the baseline, which mainly affected the V-receiver, were removed by discarding several sub-scans. 
$^\flat$ Setting 19 was added later in the program. 
$^\sharp$ Setting 16 was observed twice using a slightly different value for the systemic velocity.}
\end{table*}

\section{Observations and data reduction}

The observations we present in this paper were obtained with the Herschel Space 
Observatory \citep[][]{pilbratt2010}, using the Heterodyne Instrument for the 
Far Infrared \citep[HIFI,][]{degraauw2010}. This data set is part of the 
results obtained by 
the Guaranteed Time Key Program HIFISTARS, 
which is devoted to the study of the warm gas and water vapour 
contents of the molecular envelopes around evolved stars. 
The main target lines 
of the HIFISTARS project were the $J$=6--5, 10--9, and 16--15 transitions of 
\doceCO\ and \treceCO, and several lines of ortho- and para-\water\ sampling a 
wide range of line-strengths and excitation energies, including vibrationally 
excited states. In addition, some other lines were observed simultaneously, thanks 
to the large instantaneous bandwidth coverage of the HIFI receivers. 
We observed  16 different frequency settings for \vycma.  
In the naming adopted in the project, 
these spectral setups are, in order of increasing local oscillator (LO) frequency, settings\,14, 
13, 12, 17, 11, 10, 09, 08, 07, 06, 05, 04, 03, 19, 16, and 01; their 
main observational parameters are listed in Table\,\ref{tabteles}. 

The observations were all performed using the two orthogonal linearly 
polarized receivers available for each band, named H and V ({\em horizontal} and 
{\em vertical}\/) for their mutually perpendicular orientations. These receivers work 
in double side-band mode (DSB), which doubles the instantaneous sky frequency 
coverage of the HIFI instrument: 4 plus 4\,GHz 
for the superconductor-insulator-superconductor (SIS) receivers of bands 1 to 5, and 
2.6 plus 2.6\,GHz for the hot-electron bolometer (HEB) receivers 
of band 7. 
The observations were all performed in the dual-beam switching (DBS) mode. In 
this mode, the HIFI internal steering mirror chops between the source position and two
(expected to be) emission-free positions located 3\arcmin\ at either side of the science target.
The telescope alternately locates the source in either of the two chopped beams
($a$ and $b$), providing a double-difference calibration scheme, 
(ON$_a-$OFF$_b$)$-$(OFF$_a-$ON$_b$), 
which allows a more efficient cancellation of the residual baseline and optical standing 
waves in the spectra \citep[see][for additional details]{roelfsema2012}. In our program, 
the DBS procedure worked well except for band 7, where strong ripples (generated by electrical standing waves) 
are often found in the averaged spectra, especially in the case of the V-receiver. The HIFI 
data shown here were taken using the wide-band spectrometer (WBS), an acousto-optical 
spectrometer that provides full coverage of the whole instantaneous IF band in 
the two available orthogonal receivers, with an effective spectral resolution slightly 
varying across the band, with a mean value of 1.1\,MHz. This spectrometer is made of 
units with bandwidths slightly wider than 1\,GHz, and therefore four/three units per 
SIS/HEB receiver are needed to cover the full band.

The data were retrieved from the Herschel Science Archive and were reprocessed using 
a modified version of the standard HIFI pipeline using HIPE\footnote{HIPE is a joint 
development by the Herschel Science Ground Segment Consortium, consisting of ESA, the 
NASA Herschel Science Center, and the HIFI, PACS, and SPIRE consortia. Visit  
{\tt http://herschel.esac.esa.int/HIPE\_download.shtml} for additional information.}. This customized 
level-2 pipeline provides as final result individual double-difference elementary 
integrations without performing the final time-averaging, but stitching the three or four used 
WBS sub-bands together\footnote{The standard HIPE pipeline does perform this time average, but 
does not perform sub-band stitching by default, providing just a single spectrum per 
receiver and WBS unit.}. Later on, these spectra were exported to CLASS\footnote{CLASS is 
part of the GILDAS software package, developed and maintained by IRAM, LAOG/Univ. de Grenoble, 
LAB/Obs. de Bordeaux, and LERMA/Obs. de Paris. For 
more details, see {\tt http://www.iram.fr/IRAMFR/GILDAS}} 
using the hiClass tool within HIPE for a more detailed inspection and {\em flagging} 
of data with noticeable ripple residuals. Time-averaging was also performed in CLASS, as 
well as baseline removal. Finally, spectra from the H and V receivers were compared and averaged 
together, as the differences found between the two receivers were always smaller than the 
calibration uncertainties. 

In general, the data presented no problems and did not need a lot of {\em flagging}, except 
for the settings observed using band 7 (see 
Table\,\ref{tabteles}). All these settings presented severe residual ripples whose intensity 
varied from sub-scan to sub-scan. A semi-automated procedure was designed in CLASS to detect 
and remove the sub-scans in which the ripples were more severe. The application of this procedure 
normally results in the rejection of 30\% to 50\% of the non-averaged spectra, which produces a 
final spectrum slightly noisier, but with a more reliable baseline, since the standing waves are 
largely suppressed. Some instrumental
features affecting the baseline were also detected in settings\,09, 10, and 11, which are
also largely suppressed by removing the most affected sub-scans. 

The original data were oversampled to a uniform channel spacing of 0.5\,MHz, but we smoothed 
all spectra down to a velocity resolution of about 1\,\kms.
The data were re-calibrated 
into (Rayleigh-Jeans equivalent) main-beam temperatures ($T_{\rm mb}$) adopting the latest available 
values for the telescope and main beam efficiencies \citep{roelfsema2012}. In all cases we assumed a 
side-band gain ratio of one. A summary of these telescope characteristics, including the total 
observational uncertainty budget, is given in Table\,\ref{tabteles}. 

\section{HIFI results}\label{sec:results}

The final results of the HIFISTARS observations of \vycma\  are presented in Figs.\,\ref{fig-wbs-a} and
\ref{fig-wbs-b}
in Appendix\,\ref{sec:full-wbs-spectra}, where 
we show the full bandwidth observed in each of the settings using the WBS. 
The original spectral resolution was degraded down to about 
1\,\kms\ by averaging to the nearest integer number of channels, and a baseline was removed.
Because the receivers work in DSB mode, each spectral feature has two possible 
values for its rest frequency, one if the emission enters the receiver via the LSB, and another if
it does via the USB. In Figs.\,\ref{fig-wbs-a} and \ref{fig-wbs-b} we therefore plotted both frequency axes.
In all cases, frequency scales and line identifications are made assuming a systemic velocity for the 
source of 22\,\kms\ w.r.t. the \LSR\ frame. Lines from the LSB and USB, marked in red and blue, 
are labelled on the lower/upper x-axis of each panel. For the three lines that remain unidentified or unassigned, 
marked in green, we give labels for the two (LSB/USB) possible rest frequencies.

We detected 34 lines of CO and 
\water\ (including their isotopomers), 96 additional lines of 14 other species, and 3 unidentified lines. 
Therefore, in the 114 GHz covered by the observations (in both upper and lower side-bands), we 
detected a total 133 spectral lines, resulting in a mean line density of 1.17 lines per GHz. 
In Tables\,\ref{tablines1}, \ref{tablines2}, \ref{tablines3}, and \ref{tabso2}, we give the main observational parameters  of the detected lines.
We list the species and transition names, the energy of the upper 
state above the ground level of the species, the rest frequency found in the spectral line catalogues, the name 
of the HIFISTARS setting and side-band in which the line was observed, the root mean square (rms) value for the given spectral 
resolution, the peak intensity, the integrated area, and the velocity range over which the line's emission was integrated
(in the cases where this applies). For unblended well-detected 
lines, peak and area were computed directly from the observational data. For blended lines or spectral features 
for which the signal-to-noise ratio (S/N) is modest, we fitted Gaussian profiles to the spectra, and we give the 
peak and area resulting from these fits. In the case of blended lines, 
we simultaneously fitted as many Gaussian profiles --- usually two, but 
sometimes up to four --- as evident lines appeared in the blended profile. The results from these fits, 
as well as the resulting total composite profile, can be seen in the figures of the individual lines (see next paragraph).
Parameters for unidentified features are
listed in Table\,\ref{tabunknown}.

\begin{figure}
	\resizebox{\hspec}{!}{\includegraphics{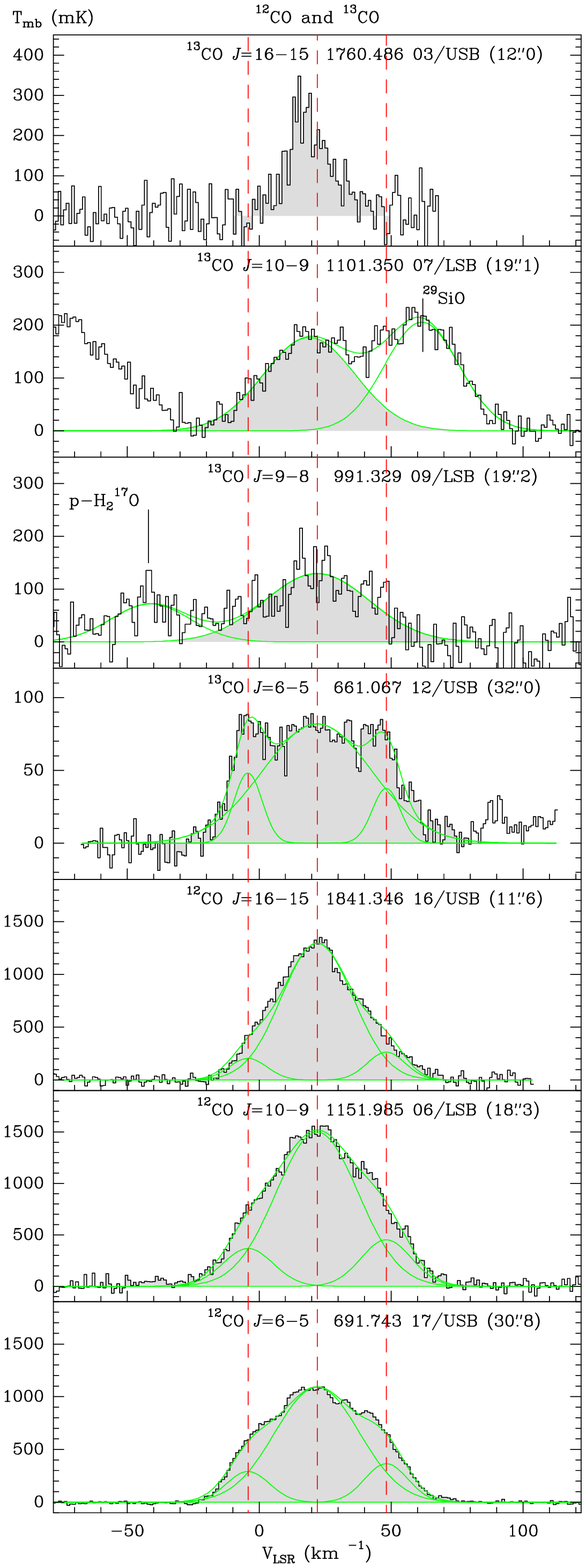}}  
	\caption{Results for \doceCO\ and \treceCO\ from HIFISTARS observations.
	Line name, rest frequency (GHz), setting and side-band in which the line has been 
	observed, and HPBW of the telescope at the frequency are also given. The
	grey area below the intensity histogram indicates the integrated line intensity 
	quoted in Table\,\ref{tablines1}. The horizontal axis gives the Doppler velocity 
	for the adopted rest frequency. The intensities are given in $T_{\rm mb}$ scale. 
	The vertical dashed lines mark the velocities adopted for the central and  
	HEVW components. The green lines show Gaussian fits either for the HEVW and main components or for
	blended lines, and their composite profile. See Sects.\,\ref{sec:results} and \,\ref{sec:CO} 
	for more details. Labels for other lines in the spectra are also given in the cases of blended lines.}
\label{fig-co}
\end{figure}

Individual plots for the different-well detected lines are
presented in Figs.\,\ref{fig-co}--\ref{fig-siovx} and \ref{fig-so}--\ref{fig-n-others}.
In these individual line plots, the horizontal scale is given in \LSR\ velocities 
assuming the rest frequency quoted in the tables. In many cases, some other lines, 
close in frequency or from the other side-band, can be seen in the plots. To avoid 
confusion, the spectral line we refer to is plotted in the centre of the panel, and 
the grey area below the histogram shows the velocity range considered when computing 
its integrated intensity as given in Tables\,\ref{tablines1}, \ref{tablines2}, \ref{tablines3}, and \ref{tabso2}. 
In the case of line blends, we also plot the results of the multi-line  
Gaussian fit we used to separate the emission from the different lines, as well
as the resulting composite spectrum. Here the grey area also shows the 
contribution to the profile of the spectral line shown in the panel. Species and line names, rest
frequency, setting and side-band, as well as the HPBW of the telescope at the corresponding
frequency are also given in each panel. 
To facilitate the comparison 
between the different line profiles, vertical red dashed lines 
mark the assumed systemic velocity and the central velocities of the two high expansion velocity components (see next 
paragraphs).

We only made observations pointed at the central position of the star: 
J2000 R.A. 07$^{\rm h}$22$^{\rm m}$58\fs33, Dec $-$25\degr46\arcmin03\farcs2 
\citep[see][]{zhang2012}. Therefore, for emitting sizes larger than or comparable 
with the beam size of the telescope, some flux is missed. In the optical and IR, 
the nebula around \vycma\ is not larger than 8\s\ in diameter (see Sect.\,\ref{sec:introduction}).  
From the interferometric maps of lines of CO and other molecules 
\citep{muller2007,verheyen2011,fu2012}, it is found that the molecular emission is also smaller 
than 8\s. In these cases, however, it can be argued that any more extended component could be 
resolved out. 
We investigated the existence of such an extended component (10\s\ in diameter or larger)
by comparing single-dish data from radio telescopes of different beam sizes. 
We compared the \doceCO\ \jdu\ spectrum from \vycma\ observed at 
the IRAM\,30m (unpublished observations by some of us) and JCMT\,15m \citep{kemper2003} telescopes,
with HPBWs of 12\s\ and 19\farcs7, respectively. When comparing these two
spectra, we found an IRAM\,30m-to-JCMT\,15m flux ratio of 2.1, except for the 
velocity range 12--22\,\kms, where we obtained values between 1.8 and 2.1.
If we assume that these ratios are only due to the different beam dilution factors, 
we derive a FWHM size for the emitting region of $\sim$\,8\s\ for the 2.1 ratio,
while the 1.8 value can be explained if the size of the source is a large as
$\sim$\,12\farcs5. We note that this analysis is affected by the uncertainties 
in the absolute calibration of both telescopes, but these sizes are compatible with 
existing images of the nebula. Still, the existence of a structure
larger than $\sim$\,8\s\ at moderate blue-shifted velocities (between 12--22\,\kms) 
agrees with the lost flux at these velocities noted by \citet{muller2007} in their interferometric observations. 
All these estimates agree with other results obtained from single-dish studies
from ground-based telescopes such as those by \citet{ziurys2007,ziurys2009}.
Our very crude size estimates should be taken as upper
limits, because they are based on a low-energy transition, with \Euppk\ of just 17\,K, 
of a very abundant and easily thermalizable molecule.  
For the observations presented here, we note that the HPBW of the
telescope is larger than these estimated sizes, and/or the excitation energies of the
transitions are considerably higher (see Tables \ref{tabteles} to \ref{tablines2}, and \ref{tablines3} to
\ref{tabunknown}). Hence no significant amount of lost flux is expected in our data.

\begin{figure*}
\sidecaption
 	\includegraphics[width=12cm]{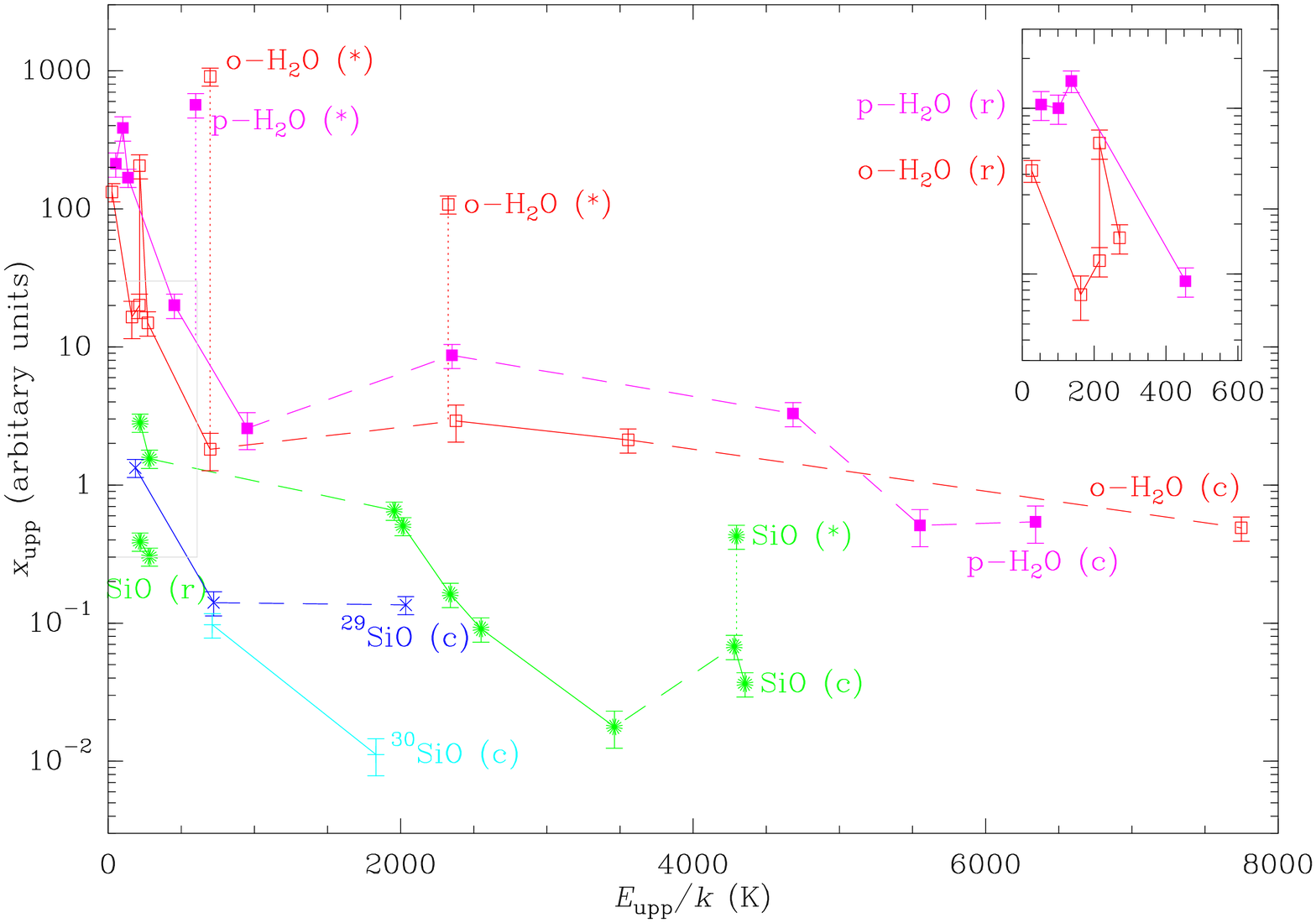} 
	\caption{Rotational diagrams for species discussed in Sect.\,\ref{sec:results} with emission from 
	vibrationally excited levels: for \water\ and \vosio, central (c) and red-HEVW (r) components; 
	for \vnsio\ and  \trsio, 
	central component (c) only. Solid lines join upper levels in
	the same vibrational state, while dashed lines join upper levels in different vibrational states.
	\vosio\ and \water\ maser lines, indicated with an asterisk (*), are connected to their respective diagrams by 
	vertical dotted lines.}
	\label{rotation-two}
\end{figure*}

\begin{figure}
 	\resizebox{\hsize}{!}{\includegraphics{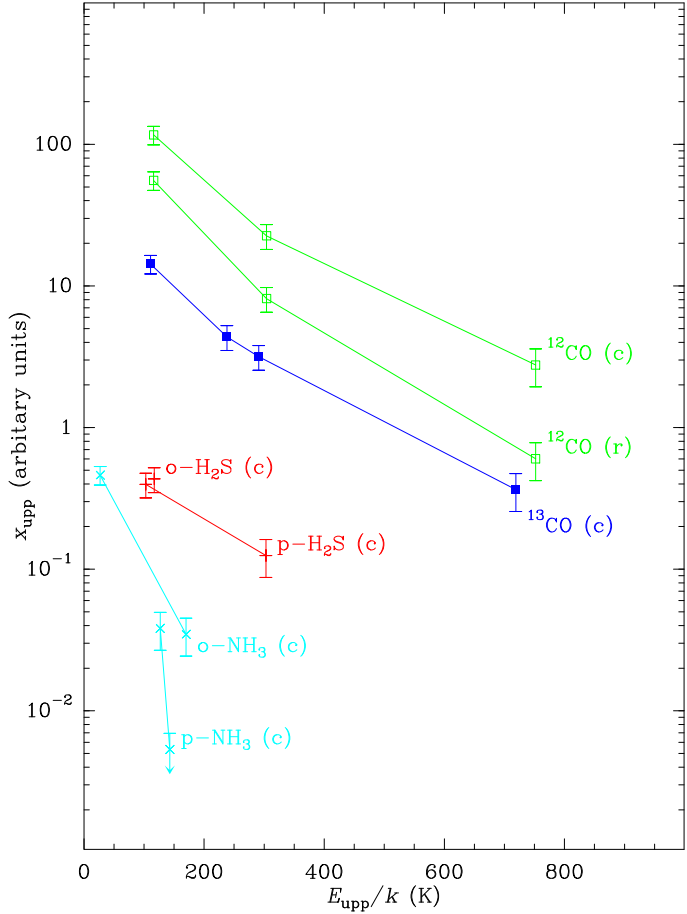}} 
	\caption{Rotational diagrams for species discussed in Sect.\,\ref{sec:results} with no emission from 
	vibrationally excited levels: for \doceCO, central (c) and red-HEVW (r) components; 
	for \treceCO, \ammonia, and \rotten, central component (c) only. Upper limits are marked with
	downward arrows.}
	\label{rotation-one}
\end{figure}

\begin{table*}
\caption{Spectral-line results for carbon monoxide and water vapour.}             
\label{tablines1}               
\begin{tabular}{rlcccccccc} 
\hline\hline 
Species and & Rotational    & $E_{\rm upp}/k^\dagger$  
                                                                     & Rest freq.       & Setting \& &  ~r.m.s.$^\ddagger$   
                                                                                                                 & Peak & Area     & Veloc. range &       \\
elec./vibr. state & quantum nums. &  (K)                                        &  (GHz)           & sideband&  (mK)        & (mK) & (K\,\kms)& \LSR\ (\kms) & Comments \\
\hline
\\
\doceCO\
$v$=0                               & $J$=~~6--~~5                   & ~~116& ~~691.473  & 17 USB &  13.4 (1.08) &  1085  &~~57.6~~ &[--50;+120]  &    \\
                                    & $J$=10--~~9                    & ~~304&  1151.985  & 06 LSB &  59.5 (1.04) &  1554  &~~80.6~~ &[--60;+100]  &    \\
                                    & \jdsq                          & ~~752&  1841.346  & 16 USB &  55.5 (0.98) &  1352  &~~52.8~~ &[--30;+80]~~ &    \\                                                                                                                                                               
$v$=1                               & $J$=~~5--~~4                   & 3166 & ~~571.021  & 14 USB & ~~6.6 (1.05) & ~$\lesssim$20 & ~$\lesssim$0.15& \\
                                    & $J$=15--14                     & 3741 &  1710.861  & 04 LSB &  66.5 (1.05) & $\lesssim$200 & ~$\lesssim$7.00&\\
\\
\treceCO\                             
          $v$=0                     & $J$=~~6--~~5                   & ~~111& ~~661.067  & 12 USB & ~~7.8 (0.91) &~~~~85  &~~~~5.02 &[--45;+80]~~ &    \\
                                    & $J$=~~9--~~8                   & ~~238& ~~991.329  & 09 LSB &  30.5 (1.06) & ~~131  &~~~~6.49 & g-fitted      &    \\
                                    & $J$=10--~~9                    & ~~291&  1101.350  & 07 LSB &  22.0 (1.09) & ~~173  &~~~~7.20 & g-fitted      &    \\
                                    & \jdsq                          & ~~719&  1760.486  & 03 USB &  78.3 (1.02) & ~~350  &~~~~5.19 &[--10;+50]~~ &    \\                                                                                                                                                              
\\
C$^{18}$O
$v$=0                               & $J$=~~6--~~5                   & ~~111& ~~658.553  & 12 USB & ~~7.8 (0.91) & ~$\lesssim$23 & ~$\lesssim$0.50&        & \\ 

\\
ortho-\water\
 $v$=0                            & \jkk=\jkkul{1}{1}{0}{1}{0}{1}  &~~~~27& ~~556.936  & 14 LSB & ~~6.6 (1.08) &  1169  &~~49.3~~ &[--55;+88]~~ &    \\
                                    & \jkk=\jkkul{3}{0}{3}{2}{1}{2}  & ~~163&  1716.770  & 04 USB &  64.5 (1.05) &  4685  &193.8~~  &[--30;+80]~~ &    \\
                                    & \jkk=\jkkul{3}{1}{2}{3}{0}{3}  & ~~215&  1097.365  & 08 LSB &  23.1 (1.10) &  1904  &~~85.0~~ &[--60;+115]  &    \\
                                    & \jkk=\jkkul{3}{1}{2}{2}{2}{1}  & ~~215&  1153.127  & 06 LSB &  59.5 (1.04) &  2993  &141.7~~  &[--65;+85]~~ &    \\
                                    & \jkk=\jkkul{3}{2}{1}{3}{1}{2}  & ~~271&  1162.912  & 06 USB &  59.5 (1.03) &  2140  &~~96.8~~ &[--50;+87]~~ &    \\
                                    & \jkk=\jkkul{5}{3}{2}{4}{4}{1}  & ~~698& ~~620.701  & 13 USB & ~~5.4 (0.97) &  1716  &~~39.5~~ &[--55;+120]  & maser~~~\\
                                    & \jkk=\jkkul{5}{3}{2}{5}{2}{3}  & ~~698&  1867.749  & 01 USB &  90.9 (0.96) &  1802  &~~60.2~~ & g-fitted      &    \\                                                                                                                                                              
         \nudu                      & \jkk=\jkkul{1}{1}{0}{1}{0}{1}  & 2326 & ~~658.004  & 12 USB & ~~7.8 (0.91) &  5654  &~~66.9~~ &[--30;+80]~~ & maser~~~\\
                                    & \jkk=\jkkul{2}{1}{2}{1}{0}{1}  & 2379 & 1753.914   & 03 USB &  78.3 (1.02) &  1007  &~~34.8~~ & g-fitted      &      \\
                                    & \jkk=\jkkul{8}{2}{7}{8}{3}{4}  & 3556 & ~~968.049  & 10 LSB &  19.8 (1.08) &~~~~46  &~~~~0.80 & g-fitted      & \\                                                                                                                                                              
         $v_{1,2}$=1,1            & \jkk=\jkkul{3}{1}{2}{3}{0}{3}  & 7749 & 1194.829   & 05 LSB &  88.8 (1.01) & ~~366  &~~~~2.12 &[--10:+50]~~ & maser\,?\\                                                                                                                                                               
\\
para-\water\ 
      $v$=0                       & \jkk=\jkkul{1}{1}{1}{0}{0}{0}  &~~~~53&  1113.343  & 07 USB &  22.0 (1.08) &  3141  & 136.5~~ &[--50;+100]  &    \\
                                    & \jkk=\jkkul{2}{0}{2}{1}{1}{1}  & ~~101& ~~987.923  & 09 LSB &  33.1 (1.06) &  3368  & 139.4~~ &[--30;+75]~~ &    \\
                                    & \jkk=\jkkul{2}{1}{1}{2}{0}{2}  & ~~137& ~~752.033  & 11 LSB &  13.6 (1.00) &  1637  &~~79.4~~ &[--50;+115]  &      \\
                                    & \jkk=\jkkul{4}{2}{2}{4}{1}{3}  & ~~454&  1207.639  & 05 USB &  88.8 (0.99) &  1299  &~~66.7~~ &[--60;+100]  &      \\
                                    & \jkk=\jkkul{5}{2}{4}{4}{3}{1}  & ~~599& ~~970.315  & 10 LSB &  19.8 (1.08) &  3188  &~~72.3~~ &[--30;+95]~~ & maser~~~\\
                                    & \jkk=\jkkul{6}{3}{3}{6}{2}{4}  & ~~952&  1762.043  & 03 USB &  78.3 (1.02) &  1011  &~~33.8~~ & g-fitted      &      \\                                                                                                                                                              
         \nudu                      & \jkk=\jkkul{1}{1}{1}{0}{0}{0}  & 2352 & 1205.788   & 05 USB &  88.8 (0.99) & ~~244  &~~~~7.36 & g-fitted      &    \\                                                                                                                                                               
         \nudd                      & \jkk=\jkkul{2}{1}{1}{2}{0}{2}  & 4684 & 1000.853   & 09 USB &  33.1 (1.05) & ~~114  &~~~~3.32 & g-fitted      &      \\                                                                                                                                                              
         \nuuu                      & \jkk=\jkkul{3}{2}{2}{3}{1}{3}  & 5552 & 1868.783   & 01 USB &  90.9 (0.96) & ~~217  &~~~~2.01 & g-fitted      & tent. detec. \\                                                                                                                                                               
         \nutu                      & \jkk=\jkkul{6}{3}{3}{6}{2}{4}  & 6342 & 1718.694   & 04 USB &  64.5 (1.05) & ~~169  &~~~~4.79 & g-fitted      & tent. detec.\\                                                                                                                                                           
\\
ortho-\waterdo\
$v$=0                             & \jkk=\jkkul{3}{1}{2}{3}{0}{3}  &~~215 &  1095.627  & 08 LSB &  23.1 (1.10) & ~~229  &~~~~5.68 &[--10;+100] &     \\                                                                                                                                                               
\\
para-\waterdo\
$v$=0                             & \jkk=\jkkul{1}{1}{1}{0}{0}{0}  &~~~~53&  1101.698  & 07 LSB &  22.0 (1.09) & ~~161  &~~~~3.53 & g-fitted     &    \\                                                                                                                                                               
\\
ortho-\waterds\
$v$=0                             & \jkk=\jkkul{3}{0}{3}{2}{1}{2}  &~~162 &  1718.119  & 04 USB &  64.5 (1.05) & ~~350  &~~~~9.59 & g-fitted     &     \\
                                  & \jkk=\jkkul{3}{1}{2}{3}{0}{3}  &~~215 &  1096.414  & 08 LSB &  23.1 (1.10) & ~~281  &~~~~7.59 &~~[+0;+60]~~ &    \\                                                                                                                                                
\\
para-\waterds\
$v$=0                             & \jkk=\jkkul{1}{1}{1}{0}{0}{0}  &~~~~53&  1107.167  & 08 USB &  23.1 (1.08) & ~~169  &~~~~4.16 &[--20;+60]~~ &     \\
                                  & \jkk=\jkkul{2}{0}{2}{1}{1}{1}  &~~101 & ~~991.520  & 09 LSB &  33.1 (1.06) &~~~~72  &~~~~2.72 & g-fitted     & tent. detec.\\
\\
HDO
     $v$=0                        & \jkk=\jkkul{3}{1}{2}{3}{0}{3}  &~~168 & ~~753.411  & 11 LSB &  13.6 (1.00) &$\lesssim$40 &$\lesssim$1.00&              & \\ 
\\
\hline
\end{tabular}
\tablefoot{
$^\dagger$ Since the ortho and para spin isomer variants of water behave as different chemical species, for all ortho-water 
isotopomers the level excitation energies are relative to their lowest (ortho) energy level, the \jkk=\jkkl{1}{0}{1}, while for the 
para species they are given w.r.t. the \jkk=\jkkl{0}{0}{0} level.   
$^\ddagger$ For the spectral resolution in \kms\ given in parentheses.}
\end{table*}

\begin{table*}
\caption{Spectral-line results for ammonia, hydrogen sulphide, hydroxyl, and silicon monoxide.}             
\label{tablines2}              
\begin{tabular}{rlcccccccc}   
\hline\hline 
Species and &  Rotational           & $E_{\rm upp}/k^\dagger$  
                                                                     & Rest freq.       & Setting \& &  ~r.m.s.$^\ddagger$   
                                                                                                                 & Peak & Area     & Veloc. range &       \\
elec./vibr. state & quantum nums. &  (K)                                        &  (GHz)           & sideband&  (mK)        & (mK) & (K\,\kms)& \LSR\ (\kms) & Comments \\
\hline
\\
ortho-\ammonia\
$v$=0                             & \jk=\nhtuz--\nhtzz             &~~~~27& ~~572.498  & 14 USB & ~~6.6 (1.05) & ~~308  &~~15.5~~ & [--50;+65]~~&    \\
                                    & \jk=\nhttz--\nhtdz             & ~~170&  1763.525  & 19 LSB &  77.8 (1.02) &  1441  &~~68.9~~  &  g-fitted    &      \\
\\
para-\ammonia\ 
$v$=0                             & \jk=\nhttd--\nhtdd             & ~~127&  1763.821  & 19 LSB &  77.8 (1.02) & ~~450  &~~21.0~~  & g-fitted     &      \\
                                    & \jk=\nhttu--\nhtdu             & ~~143&  1763.602  & 19 LSB &  77.8 (1.02) &$\lesssim$100&$\lesssim$4.5~~& &  \\                                                                                                                                                               
\\
ortho-\rotten\
$v$=0                             & \jkk=\jkkul{3}{1}{2}{2}{2}{1}  & ~~117&  1196.012  & 05 LSB &  88.8 (1.01) & ~~177  &~~10.3~~ & g-fitted     & \\
\\
para-\rotten\
$v$=0                             & \jkk=\jkkul{3}{1}{3}{2}{0}{2}  & ~~103&  1002.779  & 09 USB &  33.1 (1.05) &~~~~97  &~~~~4.22  & g-fitted    & tent. detec.\\
                                    & \jkk=\jkkul{5}{2}{4}{5}{1}{3}  & ~~303&  1862.436  & 01 LSB &  90.9 (0.97) & ~~590  &~~11.8~~ & g-fitted     & \\                                                                                                                                                
\\
\hydroxyl\
\twopihalf\ $v$=0                   & \jtmdm                         &~~270 &  1834.747  & 16 LSB &  55.5 (0.98) &  1768  &~~85.5~~  & [--40;+90]~~& \\
\\
\vosio\
    $v$=0                           & $J$=14--13                     &~~219 & ~~607.608  & 13 LSB & ~~5.4 (0.99) & ~~370  &~~15.5~~  & [--40;+105] & \\
                                    & $J$=16--15                     &~~283 & ~~694.294  & 17 USB &  13.4 (1.08) & ~~384  &~~15.1~~  & [--35;+85]~~& \\                                                                                                                                                               
    $v$=1                           & $J$=13--12                     & 1957 & ~~560.325  & 14 LSB & ~~6.6 (1.08) &~~~~76  &~~~~2.13  & [--15;+67]~~& \\
                                    & $J$=15--14                     & 2017 & ~~646.429  & 12 LSB & ~~7.8 (0.93) & ~~108  &~~~~3.01  & [--20;+60]~~& \\
                                    & $J$=23--22                     & 2340 & ~~990.355  & 09 LSB &  33.1 (1.06) & ~~178  &~~~~5.12  & g-fitted    & \\
                                    & $J$=27--26                     & 2550 &  1161.945  & 06 USB &  59.5 (1.03) & ~~190  &~~~~5.36  & g-fitted    & \\
                                    & $J$=40--39                     & 3462 &  1717.236  & 04 USB &  64.5 (1.05) & ~~160  &~~~~4.98  & g-fitted    & tent. detec.\\                                                                                                                                                             
    $v$=2                           & $J$=26--25                     & 4282 &  1111.235  & 07 USB &  22.0 (1.08) & ~~140  &~~~~3.43  & g-fitted    & \\
                                    & $J$=27--26                     & 4297 &  1153.804  & 06 LSB &  59.5 (1.04) & ~~925  & ~~25.3~~ & g-fitted    & maser\,?\\
                                    & $J$=28--27                     & 4355 &  1196.353  & 05 LSB &  88.8 (1.01) &~~~~50  &~~~~2.50  & g-fitted    & tent. detec. \\                                                                                                                                                               
\\
\vnsio\  
$v$=0                               & $J$=13--12                     &~~187 & ~~557.179  & 14 LSB & ~~6.6 (1.08) & ~~138  &~~~~5.73  & [--15;+75]~~& \\
                                    & $J$=26--25                     &~~722 &  1112.798  & 07 USB &  22.0 (1.08) & ~~203  &~~~~7.07  & g-fitted    & \\                                                                                                                                                               
        $v$=1                       & $J$=16--15                     & 2036 & ~~680.865  & 17 LSB &  13.4 (1.10) &~~~~35  &~~~~0.97  & g-fitted    & \\
\\
\trsio\     
  $v$=0                             & $J$=26--25                     &~~713 &  1099.708  & 07 LSB &  22.0 (1.09) & ~~231  &~~~~6.47  & [--20;+70]~~& \\
                                    & $J$=42--41                     & 1832 &  1771.251  & 19 USB &  77.8 (1.02) & ~~153  &~~~~3.68  & g-fitted    & tent. detec.\\
\\
\hline
\end{tabular}
\tablefoot{
$^\dagger$ 
Since ortho- and para-\ammonia\ behave as different chemical species, for para-\ammonia\ the level excitation energies are relative 
to that of the lowest para level (\jk=$1^{-}_{1}$), while  for ortho-\ammonia\ are given w.r.t. the \jk=\nhtzz\ level. 
Since the ortho and para spin isomer variants of hydrogen sulphide behave as different chemical species, for ortho-\rotten\ 
the level excitation energy is relative to its lowest (ortho) energy level, the \jkk=\jkkl{1}{0}{1}, while for the 
para species they are given w.r.t. the \jkk=\jkkl{0}{0}{0} ground level.
$^\ddagger$ For the spectral resolution in \kms\ given in parentheses. }
\end{table*}

Because the envelope around \vycma\ is peculiar (it is not dominated by a constant 
mass-loss wind that isotropically expands at constant velocity), the observed line profiles differ
from those typical of circumstellar envelopes. 
\vycma's molecular lines are very often characterized by 
up to three components: a central component
close to the systemic velocity of 22\,\kms\ and two high expansion velocity winds (HEVWs)
placed symmetrically w.r.t. the central one at 
$\sim$~22$\pm$26.2\,\kms, that is, approximately at $-$4.2 and 48.2\,\kms.
This profile decomposition was first noted by \citet{muller2007} and \citet{ziurys2007} 
and has been systematically used for
the analysis of the spectral lines of \vycma\ since, although some authors have identified up to four components in
some cases \citep{fu2012}.  
We refer to these three components as the blue-HEVW, the central or
systemic component, and the red-HEVW. 
Some lines exhibit all three
components, others only the central one, and in some others the emission in the 
HEVWs is much stronger than that of the central component. No lines match this latter case in our survey, 
but some good examples can be found in the literature, for instance, the
SO, SO$_2$, and HNC lines presented in \citet{ziurys2007}, \citet{tenenbaum2010c}, and references therein, many of which are emitted 
from levels with lower energies above the ground than the lines detected by us.

We adopted this component splitting in our work and 
tried to identify the contribution from these three components by
fitting the spectra of some lines with three Gaussian shapes at fixed {\em LSR} velocities
of $-$4.2, 22, and 48.2\,\kms. To facilitate the fitting, we also
imposed that the widths of the two HEVWs were the same.
We note that at lower frequencies, particularly in lines that are dominated
by the HEVW components like those of the species mentioned before, the width of
the red-HEVW is noticeably broader than that of the blue counterpart 
\citep[see e.g. Figs.\,1 and 3 in][]{tenenbaum2010c}. However, this seems not to be our case.
We see no indications of such a large difference between the red- and blue-HEVW widths in our spectra.
For the few cases in which our lines have a high S/N and are free from blending of other lines,
namely the three \doceCO\ lines and the \jsc\ line of \treceCO\ in Fig.\,\ref{fig-co}, the {\mbox{\jk=\nhtuz--\nhtzz}} of
\ammonia\ and the \hydroxyl\ lines in Fig.\,\ref{fig-hydrides}, and the $J$=7--6 line of HCN in Fig.\,\ref{fig-n-others},
we tried to fit our data by allowing the widths of the two HEVWs to vary independently, which resulted
in small differences, and the red-HEVW was not always the widest of the two.
Therefore,  we decided to leave just five free parameters in the fitting procedure: the intensities of the three
components, the width of the central component, and the width of the
two HEVW components. We did not use this multi-component approach  
in the cases where the S/N was low, when there was severe
line blending (as the fitting of more than three/four components would have 
become very uncertain), and when the observed line profile clearly deviated
from this triple-peaked template, as is the case of maser lines.
Separating the molecular emission in these components allowed us to investigate
their origin by studying their differing excitations. 
Some authors have argued that the HEVWs, and in general the 
high-velocity emission, arise from a wide-opening bipolar flow,
whereas the emission from the central mid-velocity component is due
to a slower expanding isotropic envelope \citep{muller2007,ziurys2007,fu2012}. However,
other authors have tried to reproduce \vycma's profiles using
fully isotropic models consisting of multiple shells with different mass-loss
rates \citep{decin2006}.

We built rotation diagrams to readily obtain estimates
for the excitation conditions of the different molecules and line components, but 
also to help in the line identification process. In these diagrams we computed
the relative population of the upper levels, $x_{\rm upp}$, assuming that the lines
are optically thin as well as a constant small size ($\ll 12\arcsec$) for the emitting region; see 
Eq. (23) in \citet{goldsmithlanger1999}. Although these assumptions are not always valid, the
resulting diagrams still provided some 
useful information on the general trend of the excitation of a species. 
We inspected these rotational diagrams for the three components separately. 
Plots of these rotational diagrams can be seen in Figs.\,\ref{rotation-two},
\ref{rotation-one}, and \ref{rotation-others}. From these diagrams,
it is clear that the rotational temperature varies with the excitation energy
of the lines (more highly excited lines tend to show higher rotational temperatures),
as is expected in a medium with a steep temperature gradient.

In the following subsections we describe the main observational results for the most 
relevant species: CO, \water, \ammonia, \rotten, and \hydroxyl, and SiO. The other detected species and
unassigned features are discussed in detail in Appendix\,\ref{sec:appendix-zooms}. 

\subsection{Carbon monoxide (CO)}\label{sec:CO}

The \doceCO\ and \treceCO\ results
are presented in Fig.\,\ref{fig-co} and Tables \ref{tablines1} and \ref{tabfits1}. 
We detected all the $v$=0 lines that we observed:
the \doceCO\ \jsc, \jdn, and \jdsq, and the \treceCO\ \jsc, \jno, \jdn, and \jdsq.
Two vibrationally excited lines of \doceCO, the \jcc\ and \jqc\ of $v$=1 were
also observed, but were not detected; the corresponding upper limits provide
no significant constraints on the excitation of these $v$=1 states.  
We did not detect the \jsc\ line of 
\CdieciochoO, which has been detected in other O-rich sources in HIFISTARS
\citep[see][]{bujarrabalHIFIppne,justtanont2012}; the obtained upper limit
for the \CdieciochoO-to-\treceCO\ \jsc\ intensity ratio is consistent with the
values for the O-rich stars detected in \CdieciochoO\ \citep[$\lesssim$\,0.1, see][]{justtanont2012}.

The \doceCO\ spectra seen by HIFI are very different from those  
at lower frequencies and excitation-energies \citep[$J$=3--2 and below, see e.g.][]{kemper2003,elvire2010}. 
Although we were able to identify the
three {\it classical components}\/ in the spectra, even in the \jsc\ line the 
emission is highly dominated by the central component. This pre-eminence increases 
as we move up the rotational ladder: while in the \jsc\ line the central-to-HEVW 
peak ratio is about 3, and the two HEVWs represent 24\% of the total emission, 
in the \jdsq\ transition this peak ratio is about 5, and the HEVW emission only accounts for 
16\% of the total emission. The only spectrum in which the triple-peaked shape
is clearly seen is the \treceCO\ \jsc, where the central-to-HEVW peak ratio
is only 2. In fact, this spectrum resembles very much those of the \jdu\ and \jtd\ \doceCO\
lines. Although we did not try to separate the emission
from the three components in the other \treceCO\ spectra (because the \jno\ and \jdn\ 
lines are blended and the \jdsq\ line is noisy), the contribution of the HEVWs
in these lines is minor. In summary, it is evident that the excitation of CO
in the HEVWs is lower than in the central component, especially for 
\treceCO.   

In spite of the different opacities expected for the two CO isotopic substitutions, 
the rotational temperature diagrams (see Fig.\,\ref{rotation-one}) give a similar excitation for the two species.
Data for the main central component yield values between 120 and 210\,K for \doceCO, 
and just 10\,K less for \treceCO. As expected, when we examine the rotational diagram for 
the HEVW components of \doceCO, we derive lower (but not very different) temperatures, 
90 to 170\,K. 
All the CO profiles are quite symmetric; for example, the blue-to-red HEVW intensity 
ratio is between 0.77 and 0.81 for \doceCO, while for 
\treceCO\ we obtain a value of 1.29. 
This means that the opacities of the CO lines in the HEVW components cannot be 
very high, since otherwise we should have detected some self-absorption in the red part of the
spectra, because the 
temperature and velocity gradients in the envelope.
 
\subsection{Water vapour (\water)}\label{sec:H2O}

\begin{figure}
	\resizebox{\hspec}{!}{\includegraphics{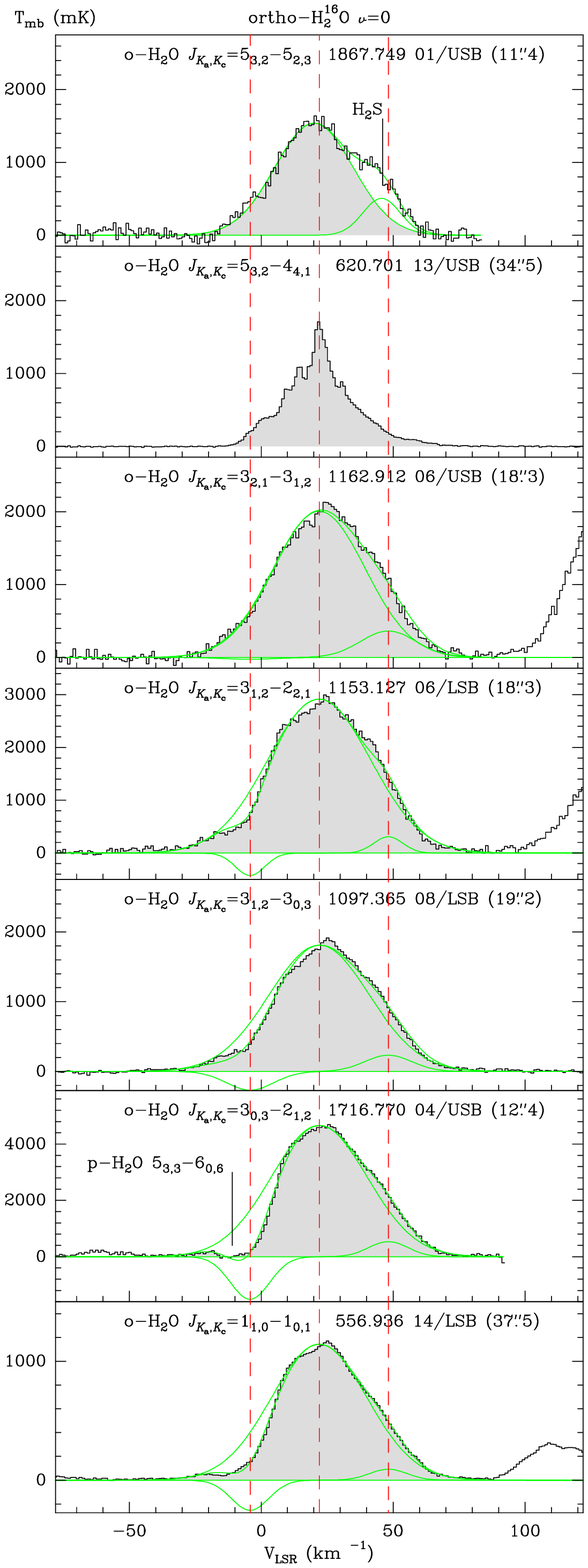}} 
	\caption{Same as Fig.\,\ref{fig-co} for the detected $v$=0 lines of ortho-\water. 
	Note the distinctive profile of the 
	\jkk=\jkkul{5}{3}{2}{4}{4}{1} maser line. The location of the undetected 
	\jkk=\jkkul{5}{3}{3}{6}{0}{6} line of \pwater\ is also indicated.}
	\label{fig-o-h2o}
\end{figure}

\begin{figure}
	\resizebox{\hspec}{!}{\includegraphics{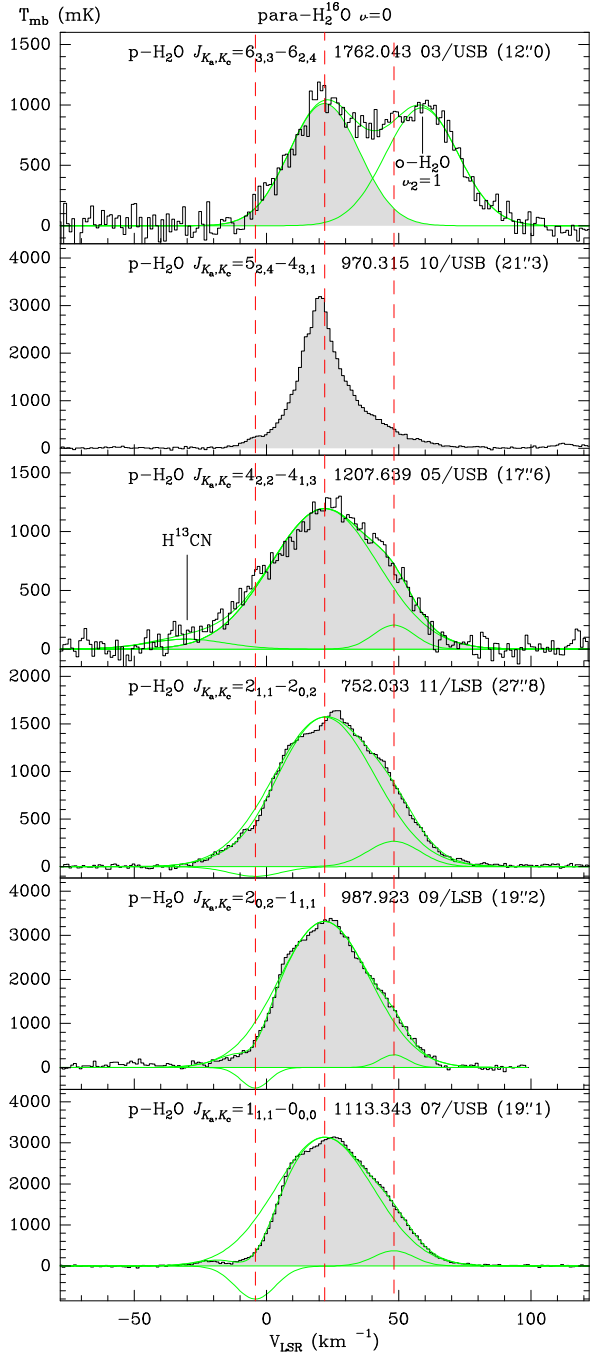}} 
	\caption{Same as Fig.\,\ref{fig-co} for the detected $v$=0 lines of para-\water. 
	Note the distinctive profile of the 
	\jkk=\jkkul{5}{2}{4}{4}{3}{1} maser line.}
	\label{fig-p-h2o}
\end{figure}

\begin{figure*}
	\resizebox{\hdoublespec}{!}{\includegraphics{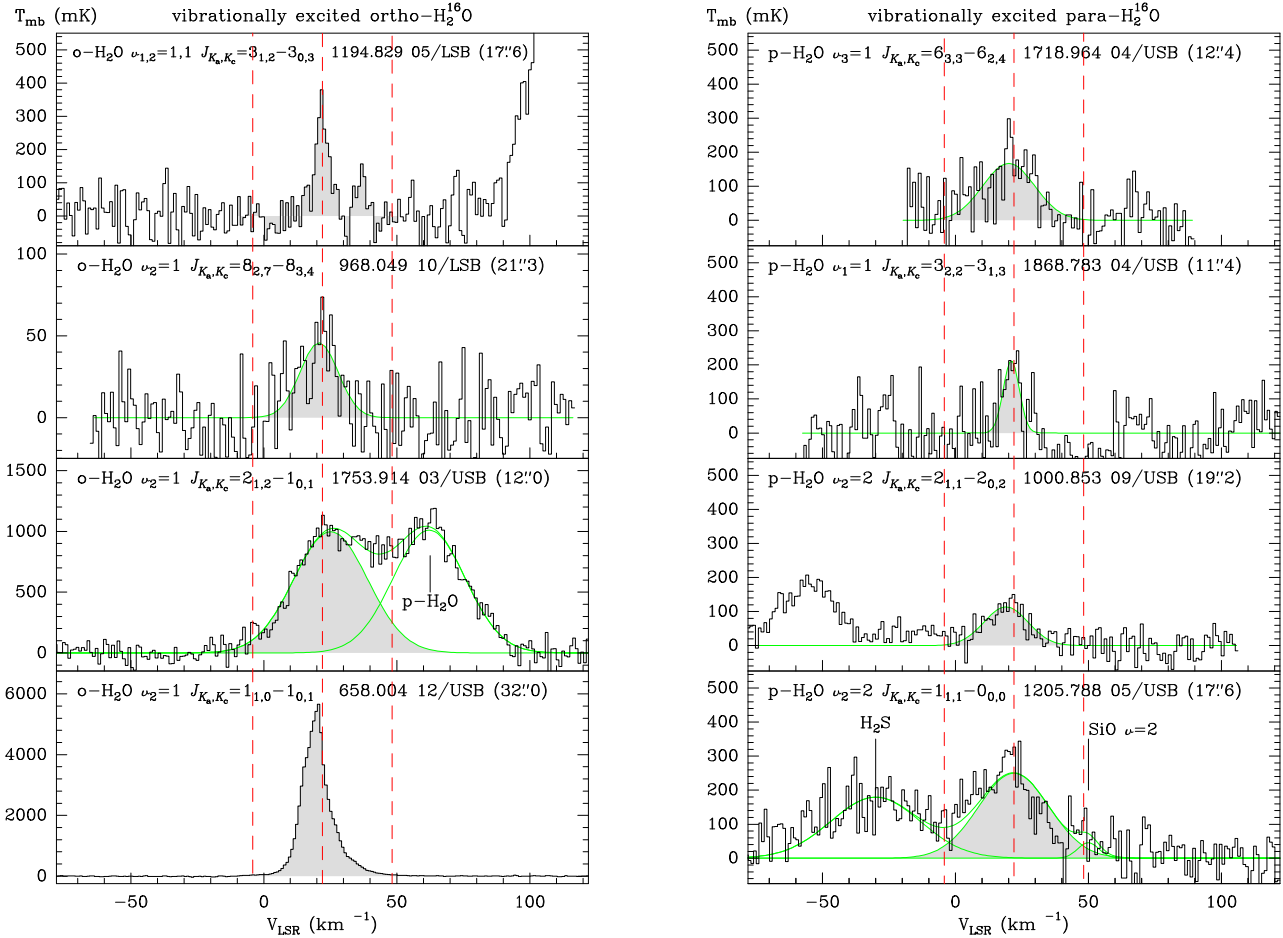}} 
	\caption{Same as Fig.\,\ref{fig-co} for the detected lines of vibrationally excited water.
	Note the narrow profile of the $v_2$=1 \jkk=\jkkul{1}{1}{0}{1}{0}{1} line in the bottom-left panel, indicating its maser origin.
	This may also be the case of the $v_{1,2}$=1,1 \jkk=\jkkul{3}{1}{2}{3}{0}{3} line in the top-left panel.}
	\label{fig-vibwater}
\end{figure*}

\begin{figure}
	\resizebox{\hspec}{!}{\includegraphics{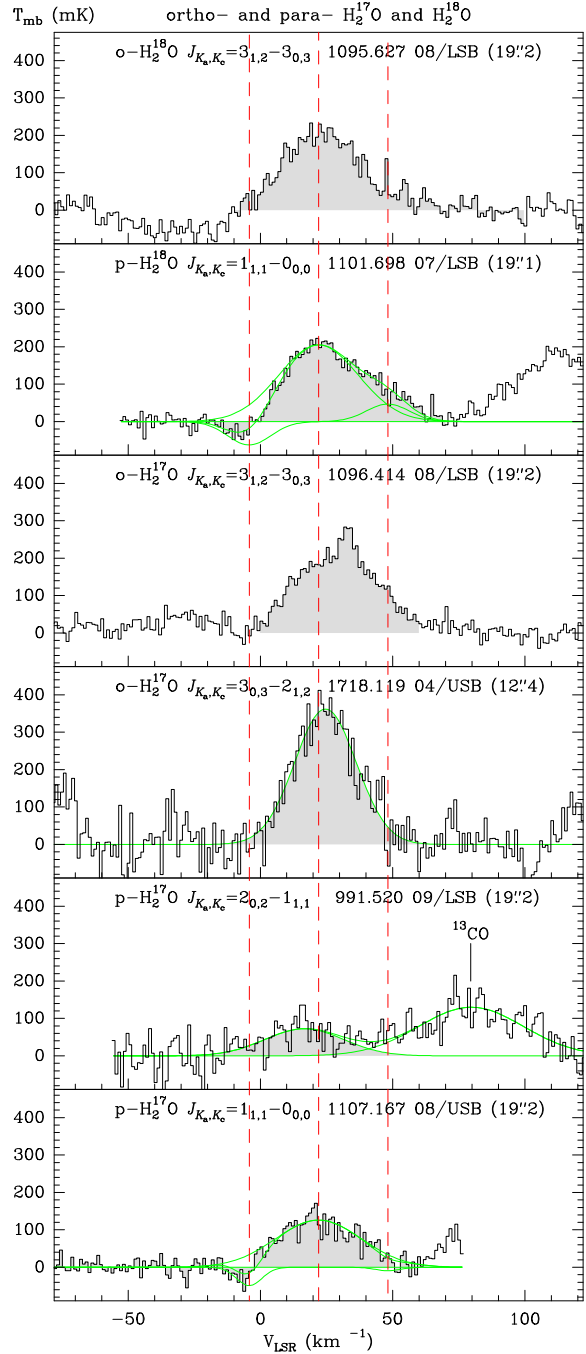}} 
	\caption{Same as Fig.\,\ref{fig-co} for the detected lines of isotopomers of water.}
	\label{fig-h2-1xo}
\end{figure}

\vycma\ HIFISTARS results for water vapour are shown in Figs.\,\ref{fig-o-h2o} to 
\ref{fig-h2-1xo}, and in Tables \ref{tablines1} and \ref{tabfits1}. In total we  
have detected 27 water lines. Seven lines of \owater, six lines of \pwater, two lines of both \owaterds\ 
and  \pwaterds, and one line of \owaterdo\ and  \pwaterdo\ each, all from the ground-vibrational 
state, and eight lines of \water\ from several vibrationally excited states. 
Frequencies of several HDO transitions were also within the observed 
bands but, as expected, no HDO line was detected.
We note the upper limit obtained the for \jkk=\jkkul{3}{1}{2}{3}{0}{3} of HDO, which is about five to seven times lower than the
intensities measured for the same transition of \waterds\ and \waterdo\ (see Table\,\ref{tablines1}).

For ortho- and para-\water, we detected all the observed transitions with high line-strengths 
($\Delta J$=0,$\pm$1 with $\Delta K_{\rm a},\Delta K_{\rm c}$=$\pm$1)
from the ground-vibrational state. We also detected the intrinsically weaker ($\Delta K_{\rm c}$=$\pm$3) 
\jkkul{5}{2}{4}{4}{3}{1} transition of \pwater, but its intensity and profile show that it is a maser
(Figs.\,\ref{fig-p-h2o} and \ref{rotation-two}), 
as is normally the case for the other low line-strength $v$=0 transitions of water detected so far. 
We also detected 
maser emission from the (intrinsically strong) \jkk=\jkkul{5}{3}{2}{4}{4}{1} transition of \owater\ 
\citep{harwit2010}. The highest excitation energy for these ground-vibrational transitions 
is the nearly 1000\,K of the upper level of the \jkk=\jkkul{6}{3}{3}{6}{2}{4} line of \pwater\ 
(see Table.\,\ref{tablines1}). We note that we did not detect the \jkk=\jkkul{5}{3}{3}{6}{0}{6}
line of para-\water, which has been observed in other O-rich envelopes in HIFISTARS 
\citep{justtanont2012}, the reason being that for \vycma\ this transition blends with the much 
stronger \jkk=\jkkul{3}{0}{3}{2}{1}{2} line due to the broad linewidths 
(see Fig.\,\ref{fig-o-h2o}).

The profiles of the lines of ortho- and para-\water, shown in Figs.\,\ref{fig-o-h2o} and
\ref{fig-p-h2o}, are not triple-peaked, but they can be decomposed into the same three
components. We did not attempt this decomposition for the
\jkkul{5}{3}{2}{4}{4}{1} (at 620\,GHz) and \jkkul{5}{2}{4}{4}{3}{1} (at 970\,GHz) lines because 
they are masers, nor for
the \jkkul{5}{3}{2}{5}{2}{3} and \jkkul{6}{3}{3}{6}{2}{4} lines because they are blended.
As for CO, the importance of the HEVW components decreases with the excitation of the line. 
The blue-HEVW component is always found in absorption, whereas the red one is 
always found in emission. This asymmetry reflects the high opacity of the water lines in general, 
and that the \water\ excitation in the HEVWs is significantly lower than in the 
central component. If we inspect the rotational diagrams for the central component of the two spin 
isomers (Fig.\,\ref{rotation-two}), we realize that interpreting this plot is more complex than in the case
of linear molecules, because the line-strength, level multiplicity, beam dilution, and frequency of the 
transitions do not monotonically increase with the excitation energy of the levels. However, we can estimate a rotation
temperature of about 200\,K for both ortho- and para-\water. Comparing the plots 
for the two spin isomers, the para-\water\ points lie about a factor 3--5 above the 
ortho points if we assume an ortho-to-para abundance ratio of 3:1. Although this value is 
affected by the probably high opacity of the lines and the
complex excitation of the molecule, the result suggests that the true ortho-to-para ratio 
must be lower than 3 \citep[see also][]{royer2010}, and hence that 
water vapour is formed under non-equilibrium chemical conditions.  
The result of inspecting the
rotational diagram for the red-HEVW component is less clear,
but suggests a temperature as low as 100\,K.

We systematically searched for a detection of water lines from its four lowest 
vibrationally excited states \nudu, \nudd, \nuuu, and \nutu\ at 2294, 4598, 5262, 
and 5404\,K above the ground; more than 30 of these lines lie within the observed frequency ranges.
We  were only able to identify the emission from seven lines, see Table\,\ref{tablines1}:
four lines from the \nudu\ state, and one line for each of the other
three vibrationally excited states listed before. We also tentatively
assigned the feature at 1194.829\,GHz (setting\,05 USB, see Fig.\,\ref{fig-wbs-b})
to the \jkkul{3}{1}{2}{3}{0}{3} line of the $v_{1,2}$=1,1 state with \Euppk\ of 7749\,K,
which holds the excitation record for all species in our survey. 
We only detected vibrationally excited lines with 
Einstein-$A$ coefficients higher than 10$^{-2}$\,s$^{-1}$, except for the \nudu\ 
\jkk=\jkkul{1}{1}{0}{1}{0}{1} and \jkkul{8}{2}{7}{7}{3}{4} lines at 658 and 968\,GHz. 
The 658\,GHz line is a maser \citep{menten-young1995}, and shows the highest peak flux in our spectral survey. 
The profile of the the \jkkul{3}{1}{2}{3}{0}{3} 
line of the $v_{1,2}$=1,1 state is also suggestive of maser amplification. The profiles and intensities of the other 
six lines are consistent with thermal emission and only display the central component at 22\,\kms.
We did not detect the \nutu\ \jkkul{3}{0}{3}{2}{1}{2} line, which was observed 
at the same time as the \nutu\ \jkkul{6}{3}{3}{6}{2}{4} and should have a similar intensity; 
for this reason and the low S/N, the detection of this latter line is considered tentative. 
The remaining non-detected vibrationally excited lines of \water\ do not provide significant constraints on the
excitation of these levels.
We note that the 658 and 970\,GHz maser lines are remarkably smooth and do not show narrow ($\sim$\,1\,\kms\ 
wide) spikes like the strong 22.2\,GHz ortho-\water\ \nuz\ \jkkul{6}{1}{6}{5}{2}{3} maser line or strong \vosio\ maser lines. 
(The 620\,GHz \nuz\ \jkkul{5}{3}{2}{4}{4}{1}, maser line shows a few narrow features.) 
In the first detection paper of the 658\,GHz line, \citet{menten-young1995} argued that the greater {\em smoothness} of this line's profile 
compared with that of the 22.2\,GHz maser, arise because the 658\,GHz maser is saturated while the 22\,GHz is 
not.

The rotational diagrams of vibrationally excited \water\ are shown in Fig.\,\ref{rotation-two}. 
Excluding maser lines, we only 
detected three excited transitions of ortho-\water\ and four lines of para-\water,
but they mostly originate from different vibrationally excited states. Therefore we can hardly investigate 
the rotational temperatures in the vibrationally excited ladders: we only derive an excitation temperature 
of 3000\,K between the two {\em thermal} \nudu\ lines of \owater. However, we can study the relative 
excitation between the different vibrationally excited states, for which we derive excitation 
temperatures between 1500 and 3000\,K. These values are much higher than
those derived for the rotational levels of the ground-vibrational state (see before), suggesting
that the population of the vibrational excited levels is mostly due to IR pumping, at 6.2 and 2.7\,\mm,
similar to the cases of other species such as SiO (sect.\,\ref{sec:SiO}) and SO (sect.\,\ref{sec:SO}).

For the rare oxygen isotopic substitutions of water, \waterds\ and \waterdo, we only detected
lines from their ground-vibrational levels with \Euppk\ below
250\,K. Several lines with excitation energies of about 500\,K  are also present in the 
observed bands, but they happen to blend with other (stronger) lines, 
and therefore it is difficult to tell whether they are detected. 
However, we did not detect the \jkkul{5}{3}{2}{4}{4}{1}
line of any of the two species, whose upper level is at more than 700\,K above the ground. 
Neither did we detect the 
\nudu\ \jkkul{1}{1}{0}{1}{0}{1} line of ortho-\waterdo. Among the six detected lines 
(see Fig.\,\ref{fig-h2-1xo}), we were able to identify the triple profile
structure only in the lowest-energy line of para-water 
(\jkkul{1}{1}{1}{0}{0}{0}) for both \waterds\ and \waterdo. As was also
the case for the main isotope, the line is very asymmetric, but here
the absorption due to the blue-HEVW is
so deep that the flux is below the continuum level, for which
we measured a value of 70--100\,mK at these frequencies. 
These P-Cygni profiles were previously observed by \citet{neufeld1999} in the main isotopomer 
in the three cases in which the line shape was resolved in their ISO observations, 
those of the \jkk=\jkkul{7}{2}{5}{6}{1}{6},  \jkkul{4}{4}{1}{3}{1}{2}, 
and \jkkul{4}{3}{2}{3}{0}{3} transitions, with $E_{\rm upp}/k$
at 1126, 702, 550\,K, respectively. 
Assuming that the
continuum level detected around 1100\,GHz has an extent similar to the
\waterds\ and \waterdo\ molecules responsible for the P-Cygni profiles, we conclude
that the opacity in the \jkk=\jkkul{1}{1}{1}{0}{0}{0} line of both species must be 
$\gsim$\,1, and that the excitation of the blue-HEVW
must be significantly lower than the true dust brightness temperature, which
is about 40\,K for a source size of 1\arcsec. 

The rotational diagrams for the central component do not give a reliable estimate of the excitation
temperature. We only have two lines per species at most, so 
we cannot distinguish between general trends and peculiarities in the
excitation of the individual lines. Taking all transitions together, we
see that the abundances of \waterdo\ and \waterds\ are very similar and 
that again an ortho-to-para abundance ratio lower than 3:1 fits the data better. 
Compared with the main isotopologue, the lines are 10 to 20 times weaker.
This is most likely due to the high opacity of the \water\ lines, but
note that the opacity of the \waterdo\ and \waterds\  cannot be negligible, as discussed above.

\subsection{Other hydrides}\label{sec:hydrides}

\begin{figure}
	\resizebox{\hspec}{!}{\includegraphics{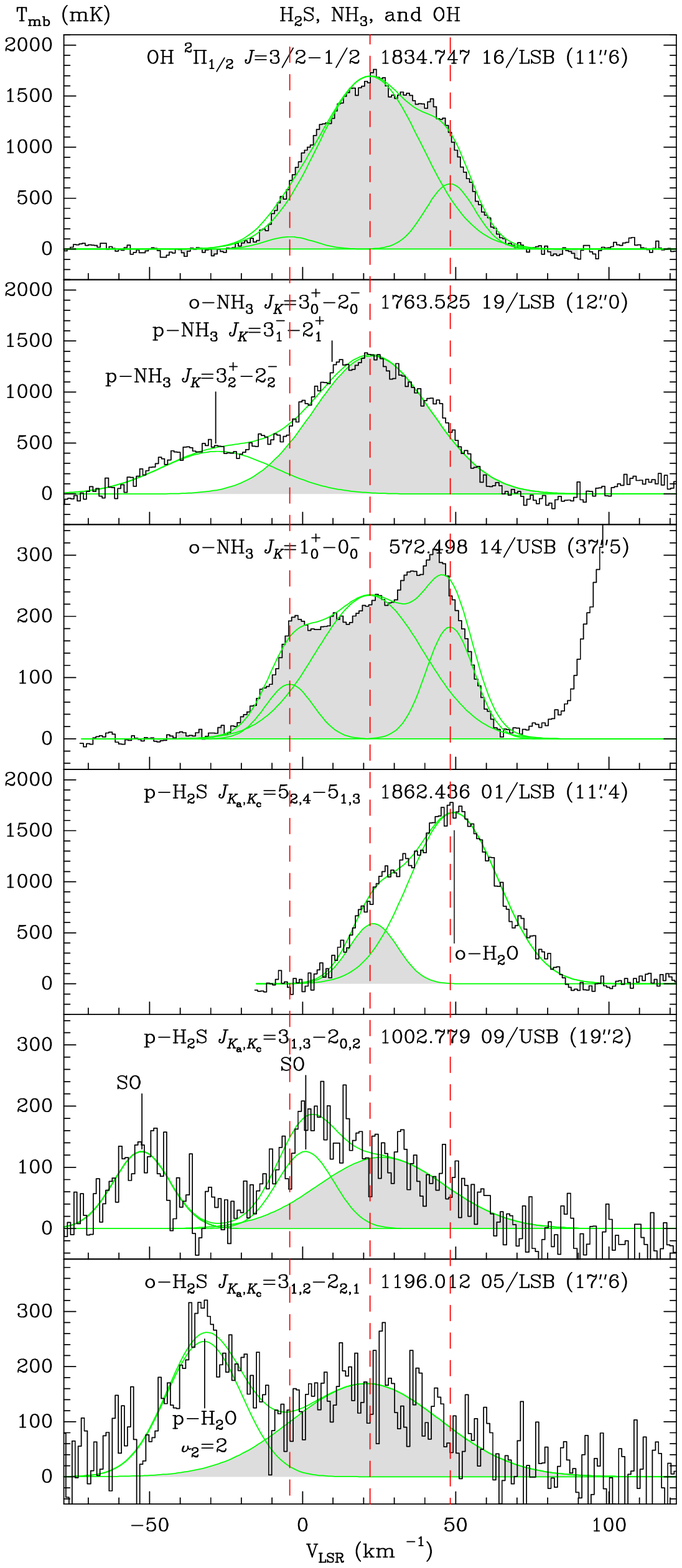}} 
	\caption{Same as Fig.\,\ref{fig-co} for the detected lines of other hydrides: \rotten.
	\ammonia, and the \hydroxyl\ radical.}
	\label{fig-hydrides}
\end{figure}  

\subsubsection{Ammonia (\ammonia)}\label{sec:NH3}

We observed four rotational lines of \ammonia, two lines of
the ortho (\jk=\nhtuz--\nhtzz\ and \nhttz--\nhtdz) and two lines of the 
para (\jk=\nhttu--\nhtdu\ and \nhttd--\nhtdd) species. These results are presented
in Fig.\,\ref{fig-hydrides}. 
The results of the \jk=\nhtuz--\nhtzz\ have already been 
published by \citet{menten2010}. The profile of this line is somehow unique in our survey
because it is  slightly U-shaped. 
The line is somewhat stronger at positive velocities, suggesting a self-absorption effect or some intrinsic
asymmetry in the emitting region, and it displays intensity bumps at 11 and 35\,\kms\
in addition to the central component at 22\,\kms and the HEVWs at $-$4 and +48\,\kms. 
This line is very similar to the \jn=\jkul{7}{6}{6}{5} of SO \citep[see Fig.\,3 in][]{tenenbaum2010c}
and resembles
those of \doceCO\ \jdu\ and \jtd\  
\citep[see][]{kemper2003,elvire2010}.
Note that the excitation energy of the  \jk=\nhtuz\ level of \ammonia, 27\,K, is in between those
of the $J$=2 and $J$=3 levels of \doceCO\ (17 and 33\,K, respectively), but is significantly lower than that of the
upper level of the \jn=\jkl{7}{6}\ of SO (47\,K). The total width\footnote{The splitting of the
hyperfine structure of these rotational lines of ammonia is very small, 
just 1.6\,\kms\ in the \juc\ line and less than 0.5\,\kms\ in the \jtd\ lines.}
of the line is also similar
to those of the CO lines, extending from about $-$30\,\kms\ to $+$70\,\kms. In spite of the
relatively more complex profile of this line, we also characterized its shape by fitting
the three Gaussians; these results are shown in 
Fig.\,\ref{fig-hydrides} and Table\,\ref{tabfits1}.

Although the three observed $J$=3--2 transitions appear to be blended, it is clear that in these higher excitation
lines only the central component is present.  We separated the contribution from 
the three lines by simultaneously fitting three Gaussians (only the central component for each
transition), assuming that the velocity
is the same
(see Fig.\,\ref{fig-hydrides}). 
Our fitting shows that both the \jk=\nhttz--\nhtdz\
and \nhttd--\nhtdd\ lines are well detected, 
showing FWHMs similar to that of the central component of the  \mbox{\jk=\nhtuz--\nhtzz} line (see Table\,\ref{tabfits1}).
In contrast, the presence of \jk=\nhttu--\nhtdu\ emission is not required by the fitting; this 
line should be at least four times weaker than the \nhttd--\nhtdd\ line. Using the two lines detected for ortho-\ammonia, 
we derive a rotational temperature of 55\,K for the 
central component, lower than what we found for other species.
This excitation value also agrees with the detected intensity 
of the \nhttd--\nhtdd\ transition if we adopt an ortho-to-para abundance ratio of 1:1, but is clearly in conflict
with the relative weakness of the \nhttu--\nhtdu\ line, whose upper level lies only 15\,K above the ground state of para-\ammonia. 
However, it must be noted that in \vycma\ the \nhttu--\nhtdu\ line of para-\ammonia\
blends with the \nhttz--\nhtdz\ one of the ortho species. 
Since the multiplicity of
the $K$=0 levels is four times that in the $K$=1 ladder, the \jk=\nhttz--\nhtdz\ line could be significantly more opaque than the 
\nhttu--\nhtdu\ one, and hence the intensity due to the para transition would be practically fully absorbed by the
ortho one if this latter is optically thick.
  
\begin{figure}
	\resizebox{\hspec}{!}{\includegraphics{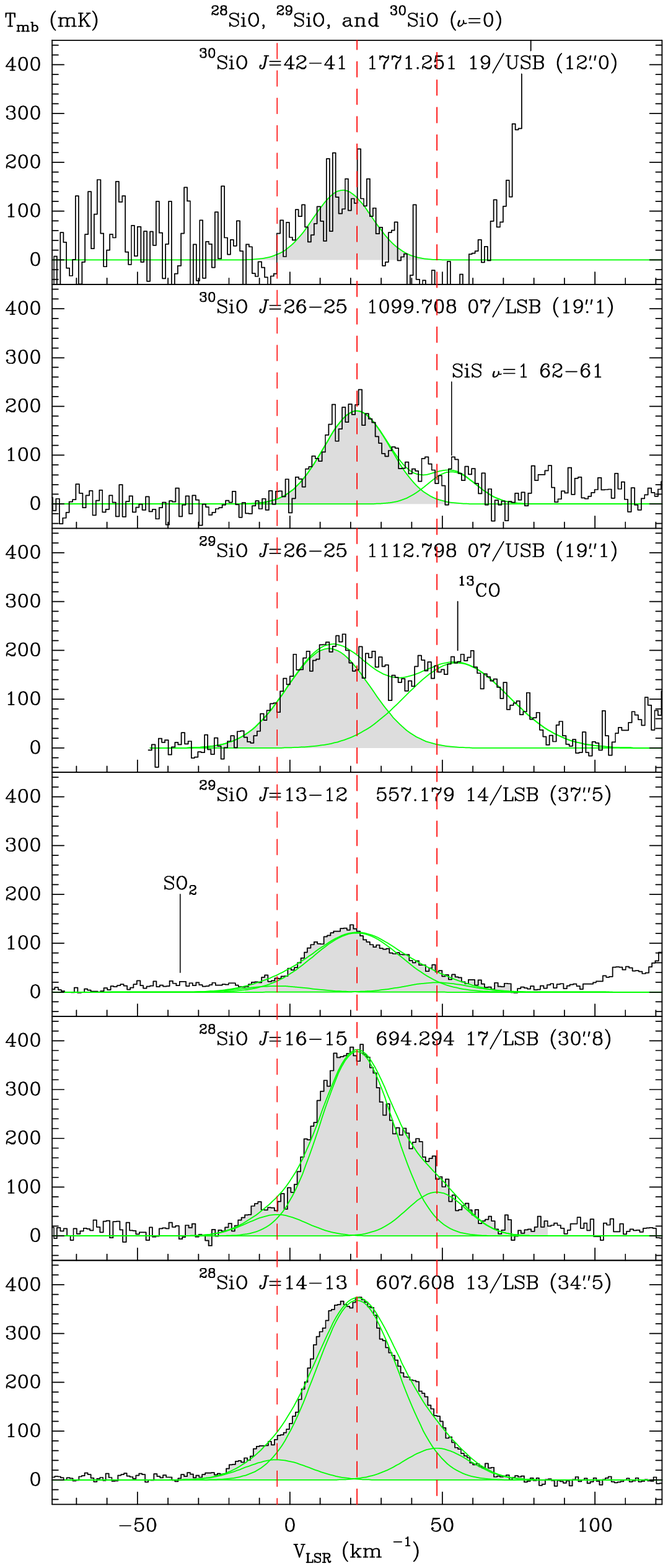}} 
	\caption{Same as Fig.\,\ref{fig-co} for the detected $v$=0 lines of \vosio, \vnsio, and \trsio.}
	\label{fig-2xsio}
\end{figure}     

\subsubsection{Hydroxyl radical and hydrogen sulphide (\hydroxyl\ and \rotten)}\label{sec:OHySH2} 

We detected the \twopihalf\ \jtmdm\ transition of OH at 
1834.747\,GHz (see Fig.\,\ref{fig-hydrides}). The line is among the strongest in our survey, 
only surpassed by the strongest lines of water. 
The profile\footnote{We note that this transition
has three hyperfine components, at $-$0.6, 0.0, $+$1.9\,\kms\ w.r.t. the quoted frequency, but this small total 
separation of 2.5\,\kms\ has little influence on the discussion that follows.} is dominated by the 
central component and displays a strong asymmetry between the two HEVWs; 
the intensity of the red-shifted component is nearly six times that of the 
blue one. The profile is among the widest we have observed in our HIFI survey, 
extending from $-$30 to $+$75\,\kms, together with the low-level lines of water.
This OH line, with \Euppk\ of 270\,K, resembles 
the \nuz\ \jkkul{3}{2}{1}{3}{1}{2} and \jkkul{4}{2}{2}{4}{1}{3} lines of \water, with \Euppk\ of 271 and 454\,K, respectively. 
The similarity between 
the profiles of lines of similar excitation of these two molecules suggests that these two 
species are also co-located, in agreement with the hypothesis that the main source of
OH formation is the photo-dissociation of \water. 

The velocity range observed in the OH 1834.747\,GHz line can be compared with those of the OH 
masers at 18\,cm, as the infrared pumping of these masers via the absorption of stellar IR photons 
at 34.6 and 53.3\,$\mu$m relates these $\Lambda$-doubling transitions in the \twopithalf\ $J$=3/2 
(ground) state with those connected by the rotational line observed with HIFI. The maser emission 
at 1612\,MHz ($F$=1$^+$--2$^-$) spans the velocity range between $-$15 and $+$60\,\kms, 
and comes from a region about 1\farcs8 in diameter \citep[see][and references therein]{benson1979}, 
3\,10$^ {16}$\,cm for a distance of 1.15\,kpc. 
The velocity range of the 1612\,MHz maser is only slightly smaller than the
1834.747\,GHz thermal line. The extreme velocities are only 10\,\kms\ lower, but
the two maxima of the \mbox{U-shaped} 1612\,MHz line occur at $\sim$ $-$8 and +50\,\kms, that is, very
close to the location of our HEVW components ($-$4.2 and 48.2\,\kms). These results
suggest that the 1612\,MHz maser arises from the shells that give rise to our HEVW
components, and that these layers have a size of just  a few arc-seconds. In addition,
the fact that the thermal line detected here is dominated by the central component
suggests that the photo-dissociation of water is taking place not only in the outermost layers 
of the envelope where the OH masers are formed, but also in the accelerated inner regions of the
circumstellar shell.

We detected 
three lines of \rotten: the \jkk=\jkkul{3}{1}{2}{2}{2}{1} of ortho-\rotten, and 
the \jkkul{3}{1}{3}{2}{0}{2} and \jkkul{5}{2}{4}{5}{1}{3} of para-\rotten. A total of 20 lines of \rotten\ lie within the HIFI observed 
bands, 
but we only detected intrinsically strong ($\Delta K_{\rm a},\Delta K_{\rm c}$=$\pm$1) transitions with \Euppk\  
$\lesssim$\,300\,K. 
The spectra of the observed lines are shown in Fig.\,\ref{fig-hydrides}. All three lines appear to be
blended with lines of SO and \water, therefore we determined their intensities by means of multiple-Gaussian 
fittings. In this case, the fittings are particularly uncertain in view of the 
resulting line parameters (widths and centroids). Therefore, the results on \rotten\ need be interpreted 
with caution. Assuming an ortho-to-para abundance ratio of 3:1, we derive a rotational temperature of 165\,K, in
agreement with the non-detection of lines with excitation energies $\gsim$\,800\,K.

\subsection{Silicon monoxide (SiO)}\label{sec:SiO}

\begin{figure*}
	\resizebox{\hdoublespec}{!}{\includegraphics{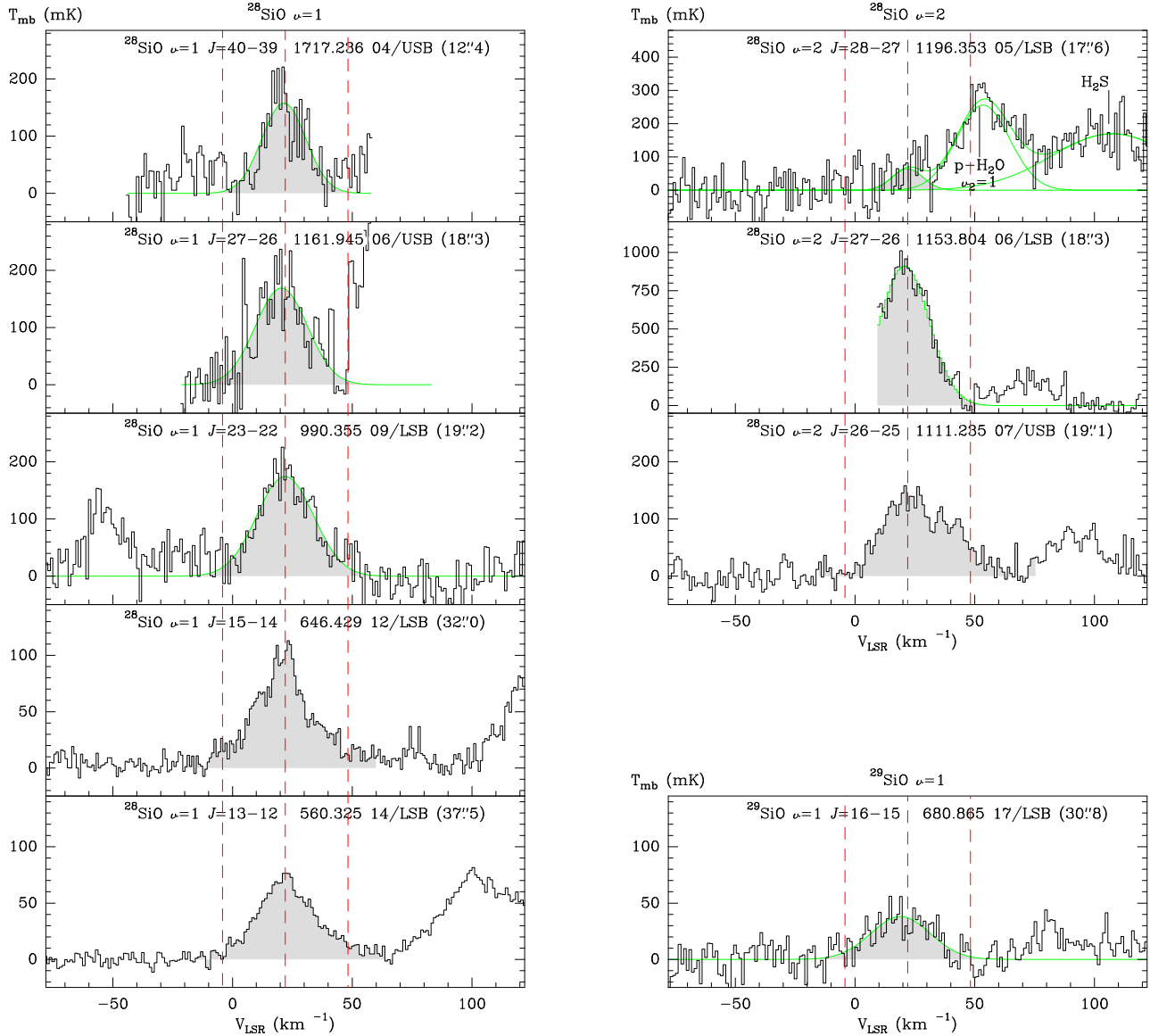}} 
	\caption{Same as Fig.\,\ref{fig-co} for the detected lines of vibrationally excited \vosio, and \vnsio\ (bottom-right panel).
    The \vosio\ $v$=2 $J$=27--26 line possibly shows some maser amplification.}	
	\label{fig-siovx}
\end{figure*}

We detected 15 lines of the three most abundant isotopic substitutions of silicon 
monoxide, \vosio, \vnsio, and \trsio, which have relative abundances of 30:1.5:1 in the Solar System. 
These results are presented in Figs.\,\ref{fig-2xsio} and \ref{fig-siovx} and in Table\,\ref{tablines2}. 
For the main isotopic substitution, \vosio, we detected the only two lines from the ground-vibrational state, 
$v$=0, that we observed, the $J$=14--13 and 16--15 transitions, with upper levels at 
excitation energies of 219 and 283\,K, respectively.
We also detected all the observed lines from the first and second vibrationally excited 
states, five $v$=1 transitions between the $J$=13--12 and the \mbox{$J$=40--39}, with $E_{\rm upp}/k$ of 
1957 and 3462\,K, and the contiguous $v$=2 transitions $J$=26--25, 27--26, and 28--27, at 
about 4300\,K above the ground level, the detection of the latter one is tentative. 
We also observed other lines of \vosio\ between rotational 
levels in vibrational 
states with yet higher excitation, $v \geq 3$, but this yielded no detections.

We detected all the observed rotational lines of \vnsio\ and \trsio\ 
from the ground $v$=0 vibrational state. These lines range from the $J$=13--12 of 
\vnsio\ at 187\,K excitation energy to the \mbox{$J$=42--41} of \trsio\ with $E_{\rm upp}/k$ of 1832\,K. 
We only detected one vibrationally excited line from these two 
rare isotopic substitutions: the 
$v$=1 \mbox{$J$=16--15 of} \vnsio. We also observed the $v$=1 \mbox{$J$=18--17} line 
of \vnsio, but this transition overlaps the strong \jkkul{2}{1}{1}{2}{0}{2}\ line of
para-\water, therefore we cannot draw conclusions on its detection. We also observed, but
failed to detect, the $v$=1 \mbox{$J$=41--40} transition of \trsio. No other $v$=1 lines of \vnsio\ or
\trsio\ are covered by our observations. We also covered several $v \geq 2$ transitions of 
the two rare isotopic substitutions of SiO, but these lines  remain undetected.  

From the comparison of the rotational diagrams for the different vibrational ladders of
\vosio, we can roughly derive the excitation conditions of 
these vibrationally excited states. This is a very relevant issue for the pumping of \vosio\  
masers from low-$J$ $v$=1 and 2 states, which are widely observed in O-rich (and some
S-type) long-period variables. From this comparison we infer vibrational excitation temperatures 
of 1500\,K between the $v$=0 and $v$=1 rotational ladders of \vosio\ and of 2500\,K 
between the $v$=1 and $v$=2 states, which are similar to the excitation energy of
these $v>0$ states (the $v$=1 and 2 states are about 1770\,K  and 3520\,K   
above the ground, respectively). These temperatures are significantly higher than those derived 
just from the comparison of $v$=1 rotational transitions of \vosio, for which we obtain 
values between 300 and 550\,K, much lower than the excitation energy of the levels, and similar
to those derived for the $v$=0 states of all SiO species.
Radiation only connects levels with $\Delta J$=1, and therefore the resulting excitation 
of the $v>0$ levels can in principle mimic the much lower collisional excitation of the 
$v$=0 states. In contrast, the collisional pumping of $v > 0$ states would yield similar  (high) excitation temperatures 
(of the order of 1700\,K and higher) for both pure rotational transitions in vibrationally excited states 
and ro-vibrational transitions. Our result supports the radiative pumping 
model for the classical \vosio\ masers ($J$=1--0 and $J$=2--1 in $v$=1 and $v$=2) against 
the collisional pumping model \citep[see][for more details]{lockett1992,bujarrabal1994}. 

The detected $v$=2 transitions of \vosio\ deserve some attention. The three lines connect
the four contiguous levels $J$=25, 26, 27, and 28. The strength of $J$=27--26 line 
is between seven and ten times higher than the other two transitions, 
suggesting some over-excitation or maser effect in this line. Although the inversion of low-$J$ masers
of \vosio\ can be explained with standard radiative models, the presence of masers in
higher-$J$ lines typically requires some additional effects, such as line overlapping with ro-vibrational lines
of \vosio\ or of some other abundant molecule (such as water). In our case, we 
identified a potential overlap between the ro-vibrational lines $J$=27--28 $v$=2--1  
of \vosio\  ($\lambda$\,=\,8.519845\,$\mu$m) and \jkk=\jkkul{10}{3}{8}{11}{4}{7}\ 
$v_2$=1--0 of ortho-\water\  ($\lambda$\,=\,8.519933\,$\mu$m), which are separated only by 3\,\kms. 
 
The line profiles of the $v$=0 lines (Fig.\,\ref{fig-2xsio}) display triangular or Gaussian-like
shapes, which are characteristic of the envelope layers where the acceleration of the gas
is still taking place and the final expansion velocity has not been attained yet. This is not
surprising because the high dipole moment of SiO favours the selection of densest gas in the envelope, 
and SiO is expected to be severely depleted onto grains as the acceleration proceeds along 
with the dust formation. However, in spite 
of this, the two detected \vosio\ transitions show a velocity range as wide as that covered by CO emission,
from about $-$30 to $+$70\,\kms (about $\pm$50\,\kms\ w.r.t. the systemic velocity of the
source). Therefore, we can conclude that there are significant amounts of SiO in the gas
phase even in the regions where the acceleration of the envelope and the grain growth 
have both ceased. A three-component Gaussian fit to the SiO lines (see Table\,\ref{tabfits1})
gives FWHMs of 28--32\,\kms\ for the central component, that is, about five to ten \kms\
narrower than those of the \doceCO\ lines with similar excitation energy; only the $J$=16--15 line
of \doceCO\ shows such a narrow central component, but at higher excitation. 
Considering
their total velocity extent and the width for single-component Gaussian fits (Fig.\,\ref{fig-siovx}), the profiles of the 
vibrationally excited SiO lines are certainly narrower than those of lines from $v$=0 states. 
\vosio\ $v$=1 lines extend from $-$10 to 60\,\kms\ at low-$J$ and from 0 to 45\,\kms\
at high-$J$. The Gaussian fits yield FWHMs of about 28\,\kms. This also applies to the the \vosio\
$v$=2 lines 
and the  $v$=1 $J$=16--15 of \vnsio. These values
are very similar to those obtained for the vibrationally excited lines of para-\water, suggesting that 
in the inner regions of the envelope where these vibrationally excited lines arise, 
the turbulence and/or acceleration are relatively strong. 

\section{Conclusions}
\subsection{Sub-mm/FIR spectrum of \vycma}

From our observations we conclude that the sub-mm/FIR spectrum of \vycma's envelope
is very rich in molecular spectral lines, in agreement with previous surveys
(see Sect.\,\ref{sec:introduction}). Considering this high 
density of lines and the broad intrinsic width of the spectral features 
in this source, between 50 and 100\,\kms\ (100\,-\,600\,MHz),  we have 
almost reached the line confusion limit. About 48\% of the observed bandwidth is occupied 
by molecular emission. In some cases, see the results from settings\,14 and 08 in Fig.\,\ref{fig-wbs-a},
it has been very difficult to find emission-free areas to use for the residual baseline 
subtraction. 
In spite of the wide frequency range covered by our survey and the large 
number of detected molecular lines, we have not identified any new species. 

We mostly detected diatomic oxides, CO, SiO, OH, SO, AlO, and PO. SiS and 
PN are the only diatomic species that do not contain oxygen. Most of the polyatomic species are 
hydrides, \water, \rotten, HCN, and the only species with four atoms, \ammonia. The 
only other polyatomic molecule is \sodos, a dioxide. 
If we include the isotopic substitutions, the spectrum is dominated by lines 
of \vosio\ (ten lines, plus five more from \vnsio\ and \trsio), ortho-\water\ (11+3) and para-\water\ 
(10+3), SO (14), HCN (8+2), CO (3+4), and PO (7). \sodos\ is a special case. We may have 
detected up to 34 lines, but some of them are indistinguishable because they are severely blended.
Of the 130 assigned lines, 24 (about 18\%) are due to vibrationally excited states. 
This relatively high prevalence of highly excited lines is most likely due to the strong IR radiation field of 
the star and the dust envelope. This efficiently pumps molecules from the $v$=0 states to  
vibrationally excited levels, greatly enhancing the emission of these transitions. We 
found that this effect is particularly strong for \water, SiO, SO, and HCN. 
The measured high intensities of some lines also require 
the contribution of an additional source of excitation, such as the line overlap of IR lines.

The structure of the lines can be decomposed into several components. The lines are typically dominated by a central
component of triangular or
Gaussian shape that peaks at the systemic velocity of the source (22\,\kms\ {\em LRS}) with FWHM values 
$\sim$\,20--45\,\kms. This emission is sometimes accompanied by two HEVW components, centred 
at $\sim$22$\pm$26\,\kms. The width of the HEVWs is narrower, between 11--25\,\kms, 
but suggests a relatively strong (either micro- or most likely large-scale)
turbulence in the gas. These two HEVW components are almost only present in lines with 
a relatively low excitation energy. Nevertheless, the extreme velocities displayed by both central and 
HEVW emission are very similar, reaching values of $\pm$50\,\kms\ w.r.t. the systemic 
velocity of the star. This high nebular expansion rate is similar to that of other super- and hyper-giants 
in the HIFISTARS sample, except for Betelgeuse, which has a significantly slower wind 
\citep[see][]{teyssier2012}.

HIFISTARS observations mostly probe the warm gas in the envelope: only \water, 
\ammonia, and \rotten\ have lines with upper levels at excitation energies below that of the $J$=6  
of \treceCO\ and \doceCO\ 
($E_{\rm upp}/k$ of 111 and 116\,K, respectively). 
In these cases we can see how the intensity of the HEVW components 
decreases with increasing excitation energy. In other species we 
mostly see the central component. Their triangular profiles suggest a strong 
acceleration in the velocity field of the gas: in some species we
detected a trend of decreasing line-width with the excitation of the line
(see Table\,\ref{tabfits1}).

The clear separation between the HEVWs and the central component suggests that the 
structures responsible for these two emissions are physically detached. The presence of 
several mutually detached shells in the envelope around \vycma\ agrees with the detailed 
images of this object (see Sect.\,\ref{sec:introduction}). These nested structures have 
also been detected in the envelopes of other massive stars \citep[\object{AFGL\,2343} and \object{IRC\,10420}, see][]{arancha2007}, 
and may be characteristic of this type of objects. We note that in \vycma\ the profiles of optically 
thin lines, such as \treceCO\ \jsc\ and \ammonia\ \jk=\nhtuz--\nhtzz, are very similar 
to other massive sources in the HIFISTARS sample \citep{teyssier2012}. All of these suggest discontinuous mass loss.
Water is detected in the HEVW and central components, and the same applies to OH, 
meaning that the formation of OH from the dissociation of \water\ does not only occur 
in the outermost layers of the envelope.  
The lower excitation and high opacity of water in the outer parts of the envelope results in 
strong self-absorbed profiles in its low-lying lines, which in the case of \waterds\ and \waterdo\ 
extend below the continuum level. 

Typically, we measured rotational temperatures between 150 and 500\,K for the central
component, even among transitions originating from vibrationally excited levels. These values 
are usually lower for the HEVW components, between 75 and 200\,K.

\subsection{\vycma\ and other massive stars in HIFISTARS}\label{sec:super-giants}

To compare our HIFISTARS results for \vycma\ with those obtained for other
hyper- and super-giants in the project, which have been published by  
\citet{teyssier2012}, we followed a procedure similar to the one
used in that paper. These authors compared the peak fluxes of the
different lines detected in \object{NML\,Cyg}, \object{IRC\,+10420}, \object{Betelgeuse}, and
\object{AFGL\,2343}, correcting for the effects of the distance, size of
the envelope, and amount of molecular gas, by normalizing all values
to the peak flux of the \jsc\ \doceCO\ line. We included \vycma\ 
in the comparison and increased the number of studied transitions. 
\begin{figure}
	\resizebox{\hsize}{!}{\includegraphics[angle=-90]{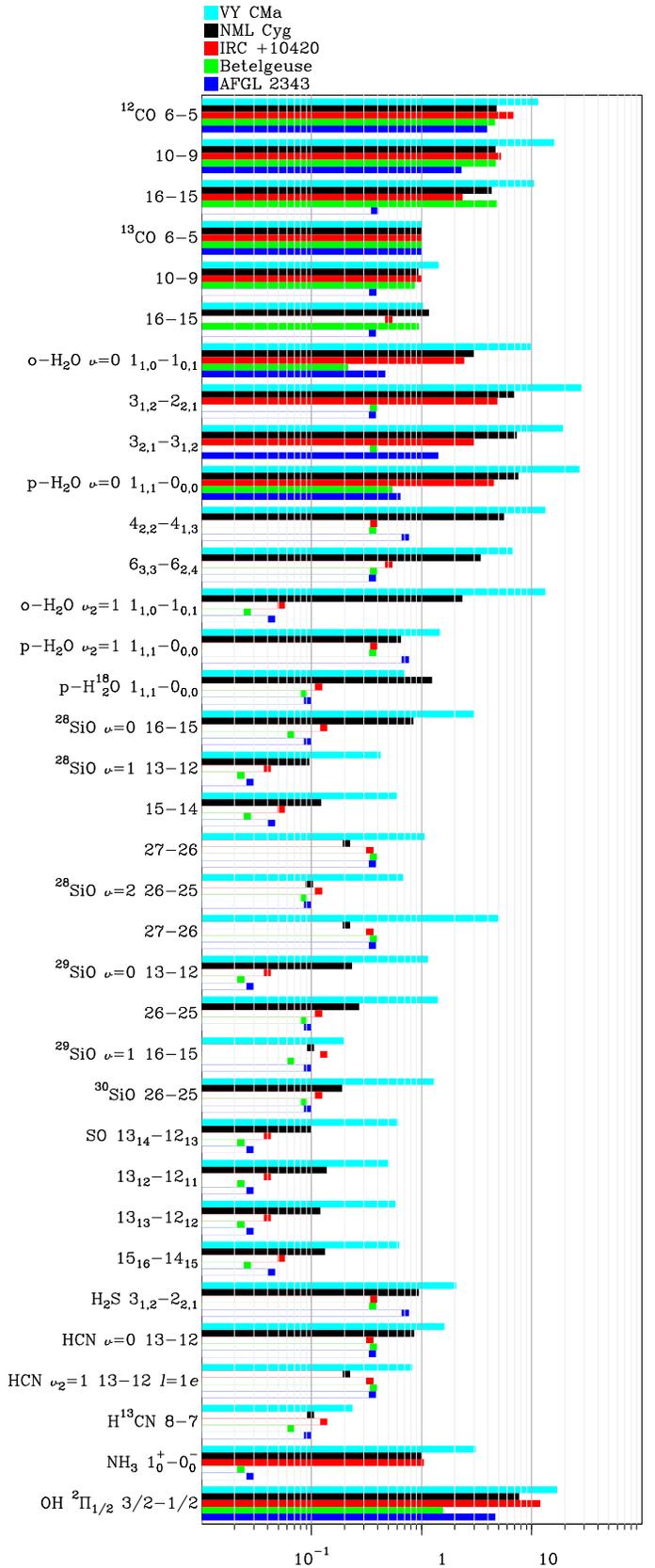}}
	\caption{Comparison of the integrated emission of different lines
	in HIFISTARS for \vycma\ and the other four hyper- and super-giants
	in the project \citep[see][]{teyssier2012}. For each object, the integrated
	intensities of the different lines have been normalized to that of its \jsc\ \treceCO\ transition.
	Horizontal bars and small squares show, respectively, 
	values and upper limits for detected and undetected lines.
	See Sect.\,\ref{sec:super-giants} for more details.}
	\label{fig-ratios}
\end{figure}
Here we chose to represent the integrated area values for the lines,
because these figures are better correlated with the total amount of molecular gas,
and, for non-detections, the corresponding upper limits provide better constraints. 
For these undetected lines, we took as upper limit the usual value of 3$\sigma\times\delta v\times\sqrt{n}$,
where $\delta v$ is the velocity resolution and $n$ is the number of channels typically covered
by the lines in the source at zero power. This corresponds to 
60\,\kms\ for \object{NML\,Cyg}, \object{IRC\,+10420}, and \object{AFGL\,2343}, and 30\,\kms\ for 
\object{Betelgeuse}. To avoid problems with the possible high opacity of the \doceCO\ lines, 
we used the integrated intensity of the \treceCO\ \jsc\ transition 
as the normalization factor. The results of this comparison are presented in Fig.\,\ref{fig-ratios}.

As we can see in this figure, \vycma\ is the strongest emitter in lines
of less abundant molecules (other than \doceCO\ and \water). This
result cannot be due to a relatively weak intensity of the \treceCO\ \jsc\
in \vycma, which might be the case if $^{13}$C were less abundant than in the
other sources, since \vycma\ is also the strongest emitter in the \treceCO\ \jdn\ and
\HtreceCN\ $J$=8--7 lines. Therefore we must conclude
that these rare molecular species are more abundant in \vycma\ than in the
other four sources. 
\vycma\ is followed by \object{NML\,Cyg}, and then by \object{IRC\,+10420}. \object{Betelgeuse} and 
\object{AFGL\,2343} 
are the weakest of all. The contrast between \vycma\ and the other sources 
is stronger for lines with higher \Euppk, such as $v > 0$ SiO lines and 
the ortho-\water\ \nudu\ \jkkul{1}{1}{0}{1}{0}{1} maser line, but also for the relatively low
excitation lines of SO included in the plot. Interestingly, the only
two lines in \vycma\ that are surpassed by \object{NML\,Cyg} are  
\treceCO\ \jdsq\ and para-\waterdo\ \jkkul{1}{1}{1}{0}{0}{0}. 
The relative weakness of the  para-\waterdo\ in \vycma, w.r.t. \object{NML\,Cyg}, contrasts with the
strength of all other \water\ lines, suggesting a lower abundance of
this isotopic substitution in our target. We also note that \CdieciochoO\
remains undetected in our \vycma\ spectra.

\begin{acknowledgements}
   HIFI has been designed and built by a consortium of institutes and university 
   departments from across Europe, Canada, and the United States under the leadership 
   of SRON (Netherlands Institute for Space Research), Groningen, The Netherlands, and 
   with major contributions from Germany, France, and the US. Consortium members are: 
   Canada: CSA, U.Waterloo; France: CESR, LAB, LERMA, IRAM; Germany: KOSMA, MPIfR, MPS; 
   Ireland, NUI Maynooth; Italy: ASI, IFSI-INAF, Osservatorio Astrofisico di Arcetri-INAF; 
   The Netherlands: SRON, TUD; Poland: CAMK, CBK; Spain: Observatorio Astronómico Nacional (IGN), 
   Centro de Astrobiología (CSIC-INTA). Sweden: Chalmers University of Technology-MC2, R
   SS \& GARD; Onsala Space Observatory; Swedish National Space Board, Stockholm 
   University-Stockholm Observatory; Switzerland: ETH Zurich, FHNW; USA: Caltech, J.P.L., NHSC.
   
   {\em HIFISTARS: The physical and chemical properties 
   of circumstellar environments around evolved stars}, P.I V. Bujarrabal, 
   is a HIFI/Herschel Guaranteed Time Key Program (KPGT\_vbujarra\_1) 
   devoted to the study of the warm gas and water vapour contents of
   the molecular envelopes around evolved stars: AGB stars, red super- 
   and hyper-giants; and their descendants: pre-planetary nebulae, 
   planetary nebulae, and yellow hyper-giants. HIFISTARS comprises 366
   observations, totalling 11,186\,min of HIFI/Herschel telescope time.
   See {\tt http://hifistars.oan.es}; and {\tt Key\_Programmes.shtml}
   and {\tt UserProvidedDataProducts.shtml} in the Herschel web portal
   ({\tt http://herschel.esac.esa.int/})
   for additional details.

   This work has been partially supported by the Spanish 
   MICINN, program CONSOLIDER INGENIO 2010, grant ``ASTROMOL'' (CSD2009-00038); and
   by the German \emph{Deutsche Forschungsgemeinschaft, DFG\/} project
   number Os~177/1--1. A portion of this research was performed at the Jet Propulsion
   Laboratory, California Institute of Technology, under contract with the National
   Aeronautics and Space Administration. RSz and MSch acknowledge support from
   grant N203 581040 of National Science Center. K.J., F.S., and H.O. acknowledge 
   funding from the Swedish National Space Board. J.C. thanks for funding from the Spanish 
   MICINN, grant AYA2009-07304.
\end{acknowledgements}

\bibliographystyle{aa}
\bibliography{vycma-bp}

\newpage

\begin{appendix}

\section{Detailed results for species not discussed in Sect.\,\ref{sec:results}}\label{sec:appendix-zooms}

\begin{figure}
 	\resizebox{\hsize}{!}{\includegraphics{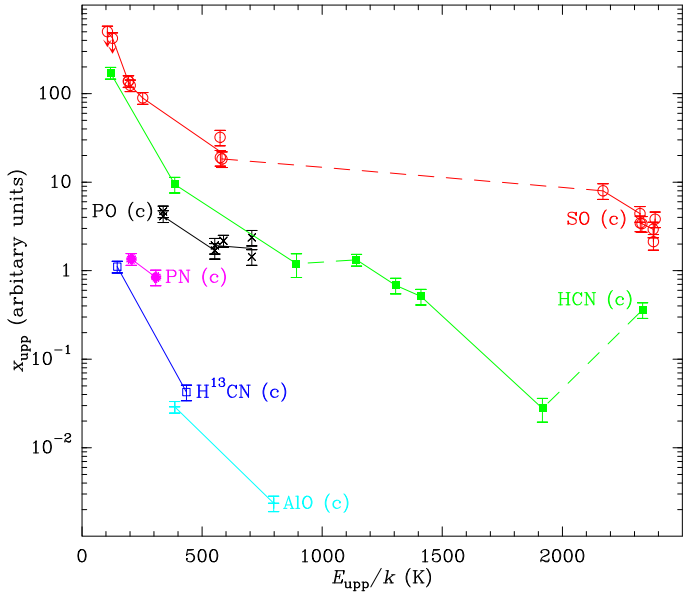}} 
	\caption{Rotational diagrams for some of the species discussed in this appendix: central component (c) only. 
	Upper limits are marked with downward arrows.
	Solid lines join upper levels in
	the same vibrational state, while dashed lines join upper levels in different vibrational states.}
	\label{rotation-others}
\end{figure} 

\subsection{Sulphur monoxide (SO)}\label{sec:SO}

We detected three rotational \jkul{J}{N}{(J{\rm-1})}{(N-1)} triplets ($N$=$J$, $J$--1 and $J$+1) 
of SO, the \mbox{$J$=13--12}, \mbox{15--14} (only one component), and \mbox{23--22}, with \Euppk\ from 
about 190 to 580\,K. We thus detected all the high line-strength SO transitions 
within the observed settings, except for one component of the \mbox{$J$=41--40}, which lies at 
one of the edges of the setting\,03 observation, and for which the upper limit obtained is not 
significant. Some other lines of SO of much  lower line-strength, with \jkul{J}{N}{(J{\rm-1})}{N} 
or with $\Delta J>1$, also lie within the observed frequency ranges but yielded non-detections. 
In general, the upper limits
of these lines are not significant either, except for those of the relatively low-lying lines
\jn=\jkul{9}{9}{8}{9} and \jkul{10}{10}{9}{10}, at 568.741 and 609.960\,GHz
respectively, and with upper-level energies of 106 and 127\,K above the ground. 
Using the information from detections and the relevant upper limits, we
built a rotation diagram for SO for the central component (see Tables\,\ref{tablines3} and \ref{tabfits1} and 
Fig.\,\ref{rotation-others}). 
We derive a rotation 
temperature
of about 200--250\,K, similar to that derived for other species from levels
of similar excitation energies. The upper limits obtained for the lines at 106 and 127\,K excitation energy
indicate that the central component of SO cannot have a significant contribution from gas
colder than about 60\,K.

\begin{figure}
	\resizebox{\hspec}{!}{\includegraphics{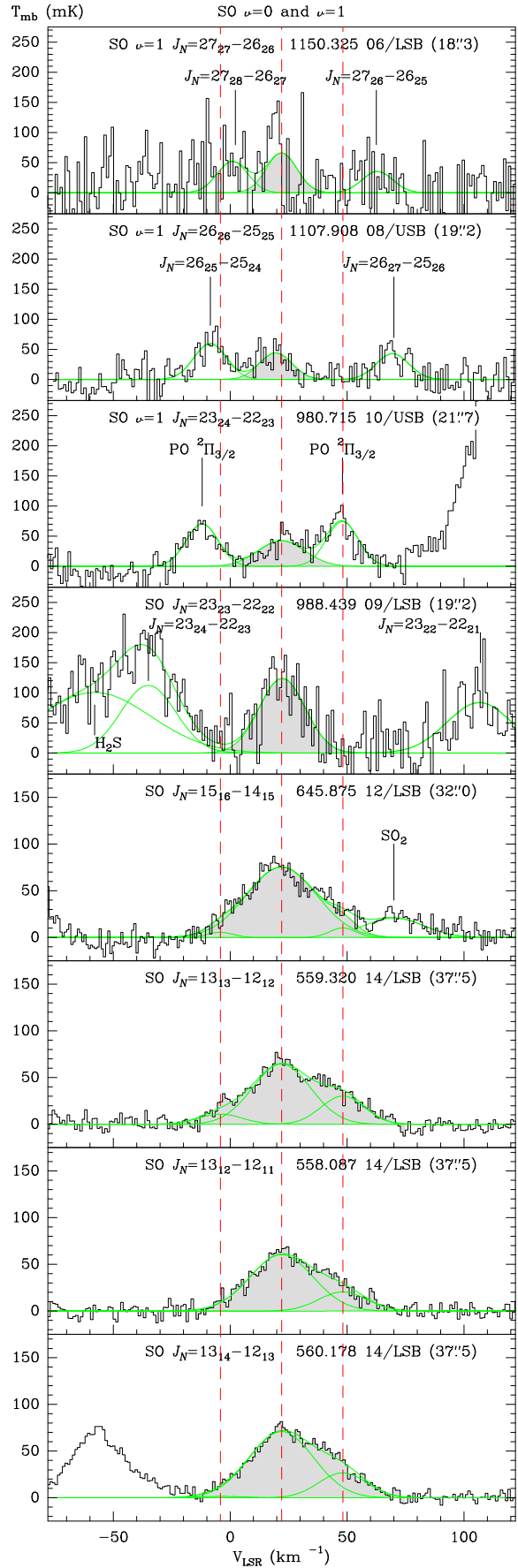}}
	\caption{Same as Fig.\,\ref{fig-co} for the detected lines of SO. The $v$=0 \mbox{$J$=23--22} and
	$v$=1 $J$=26--25 and 27--26 triplets are plotted in one panel each (fourth, second, and first from top).}
	\label{fig-so}
\end{figure}

\begin{table*}
\caption{Spectral-line results for sulphur monoxide, phosphorous monoxide, aluminium monoxide, silicon sulphide,
         hydrogen cyanide, and phosphorous nitride.}             
\label{tablines3}      
\begin{tabular}{rlcccccccc} 
\hline\hline 
Species and & Rotational    & $E_{\rm upp}/k$  
                                                                     & Rest freq.       & Setting \& &  ~r.m.s.$^\ddagger$   
                                                                                                                 & Peak & Area     & Veloc. range &       \\
elec./vibr. state & quantum nums. &  (K)                                        &  (GHz)           & sideband&  (mK)        & (mK) & (K\,\kms)& \LSR\ (\kms) & Comments \\
\hline
\\
SO
    $v$=0                           & \jn=~~\jkul{9}{9~}{~~8}{9}          &~~106 & ~~568.741  & 14 USB & ~~6.6 (1.05)   &~$\lesssim$20&$\lesssim$0.13~ \\ 
                                    & \jn=\jkul{10}{10}{~~9}{10}       &~~127 & ~~609.960  & 13 LSB & ~~5.4 (0.99)   &~$\lesssim$16&$\lesssim$0.10~  \\
                                    & \jn=\jkul{13}{14}{12}{13}      &~~193 & ~~560.178  & 14 LSB & ~~6.6 (1.08)   &~~~~82  & ~~3.08  &~~[--13;+75]~~& \\
                                    & \jn=\jkul{13}{12}{12}{11}      &~~194 & ~~558.087  & 14 LSB & ~~6.6 (1.08)   &~~~~69  & ~~2.54  &~~[--15;+75]~~& \\
                                    & \jn=\jkul{13}{13}{12}{12}      &~~201 & ~~559.320  & 14 LSB & ~~6.6 (1.08)   &~~~~77  & ~~2.92  &~~[--15;+75]~~& \\
                                    & \jn=\jkul{15}{16}{14}{15}      &~~253 & ~~645.875  & 12 LSB & ~~7.8 (0.93)   &~~~~82  & ~~3.15  & g-fitted    & \\
                                    & \jn=\jkul{23}{24}{22}{23}      &~~575 & ~~988.616  & 09 LSB &  33.1 (1.06)   & ~~156  & ~~5.21  & g-fitted     & \\
                                    & \jn=\jkul{23}{22}{22}{21}      &~~576 & ~~988.166  & 09 LSB &  33.1 (1.06)   &~~~~96  & ~~2.82  & g-fitted     & \\
                                    & \jn=\jkul{23}{23}{22}{22}      &~~583 & ~~988.439  & 09 LSB &  33.1 (1.06)   & ~~129  & ~~2.84  & g-fitted     & \\                                                                                                                                                               
    $v$=1							& \jn=\jkul{23}{24}{22}{23}      & 2169 & ~~980.715  & 10 USB &  19.8 (1.07)   &~~~~58  & ~~1.26  & g-fitted     & \\
                                    & \jn=\jkul{26}{27}{25}{26}      & 2169 &  1108.012  & 08 USB &  23.1 (1.08)   &~~~~60  & ~~1.09  & g-fitted     & \\
                                    & \jn=\jkul{26}{25}{25}{24}      & 2324 &  1107.723  & 08 USB &  23.1 (1.08)   &~~~~44  & ~~0.79  & g-fitted     & \\
                                    & \jn=\jkul{26}{26}{25}{25}      & 2331 &  1107.908  & 08 USB &  23.1 (1.08)   &~~~~45  & ~~0.81  & g-fitted     & \\
                                    & \jn=\jkul{27}{28}{26}{27}      & 2378 &  1150.409  & 06 LSB &  59.5 (1.04)   &~~~~52  & ~~0.86  & g-fitted     & \\                                 
                                    & \jn=\jkul{27}{26}{26}{25}      & 2379 &  1150.165  & 06 LSB &  59.5 (1.04)   &~~~~52  & ~~0.86  & g-fitted     & \\             
                                    & \jn=\jkul{27}{27}{26}{26}      & 2386 &  1150.325  & 06 LSB &  59.5 (1.04)   &~~~~65  & ~~1.08  & g-fitted     & \\
\\
PO 
\twopithalf\ $v$=0                  & $J$=35/2--33/2 $l$=$e$         &~~338 & ~~763.121  & 11 USB &  13.5 (0.98)   &~~~~36  & ~~1.04  & g-fitted     & \\
                                    & $J$=35/2--33/2 $l$=$f$         &~~338 & ~~763.310  & 11 USB &  13.5 (0.98)   &~~~~27  & ~~0.92  & g-fitted     & \\
                                    & $J$=45/2--43/2 $l$=$e$         &~~553 & ~~980.644  & 10 USB &  19.8 (1.07)   &~~~~78  & ~~1.06  & g-fitted     & \\
                                    & $J$=45/2--43/2 $l$=$f$         &~~553 & ~~980.834  & 10 USB &  19.8 (1.07)   &~~~~68  & ~~1.21  & g-fitted     & \\
\twopihalf\  $v$=0                  & $J$=31/2--29/2                 &~~589 & ~~680.730  & 17 LSB &  13.4 (1.10)   &~~~~15  & ~~0.62  & g-fitted     & \\                                                                                                                                                               
\twopithalf\ $v$=0                  & $J$=51/2--49/2 $l$=$e$         &~~707 &  1110.970  & 08 USB &  23.1 (1.08)   &~~~~71  & ~~2.38  & g-fitted     & \\
                                    & $J$=51/2--49/2 $l$=$f$         &~~707 &  1111.160  & 07 USB &  22.0 (1.08)   &~~~~75  & ~~1.44  & g-fitted     & \\                                                                                                                                                                 
\\
AlO
    $v$=0                           & $N$=20--19                     &~~386 & ~~764.603  & 11 USB &  13.5 (0.98)   &~~~~73  & ~~2.63  & g-fitted     & \\
                                    & $N$=29--28                     &~~798 &  1106.980  & 08 USB &  23.1 (1.08)   &~~~~84  & ~~1.00  & g-fitted     & tent. detec. \\
\\
SiS                               
    $v$=0                           & $J$=61--60                     & 1168 &  1102.029  & 07 LSB &  22.0 (1.09)   & ~~106  & ~~2.59  & g-fitted     &   \\
    $v$=1                           & $J$=55--54                     & 2404 & ~~989.685  & 09 LSB &  33.1 (1.06)   &~$\lesssim$99&$\lesssim$0.77~ \\
                                    & $J$=61--60                     & 2707 &  1096.635  & 08 LSB &  23.1 (1.10)   &~~~~44  & ~~1.19  & g-fitted     & tent. assig.  \\
                                    & $J$=62--61                     & 2760 &  1114.431  & 07 USB &  22.0 (1.08)   &~~~~63  & ~~1.61  & g-fitted     & tent. assig.  \\
                                    & $J$=64--63                     & 2870 &  1149.994  & 06 LSB &  59.5 (1.04)   &$\lesssim$179&$\lesssim$1.38~ \\
\\
HCN
    $v$=0                         & $J$=~~7--~~6                   & ~~119& ~~620.304  & 13 USB & ~~5.4 (0.97)   & ~~261  &  12.1~~ & g-fitted     & \\
                                    & $J$=13--12                     & ~~386&  1151.452  & 06 LSB &  59.5 (1.04)   & ~~266  & ~~8.11  & g-fitted     & \\
                                    & $J$=20--19                     & ~~893&  1769.877  & 19 USB &  77.8 (1.02)   & ~~145  & ~~2.85  & g-fitted     & tent. detec.\\                                                                                                                                                               
   \nudu                           & $J$=~~7--~~6 $l$=1$e$          & 1143 & ~~620.225  & 13 USB & ~~5.4 (0.97)   &~~~~23  & ~~0.85  & g-fitted     & \\
									& $J$=11--10 $l$=1$f$            & 1306 & ~~979.285  & 10 USB &  19.8 (1.07)   &~~~~64  & ~~0.90  & g-fitted     & \\
                                    & $J$=13--12 $l$=1$e$            & 1411 &  1151.297  & 06 LSB &  59.5 (1.04)   & ~~119  & ~~4.14  & g-fitted     & \\
									& $J$=20--19 $l$=1$e$            & 1917 &  1769.603  & 19 USB &  77.8 (1.02)   &~~~~65  & ~~1.22  & g-fitted     & \\ 
   \nudd                            & $J$=11--10 $l$=2$e$            & 2334 & ~~979.554  & 10 USB &  19.8 (1.07)   &~~~~31  & ~~0.45  & g-fitted     & \\                                                                                                                                                                
\\
\HtreceCN\
       $v$=0                      & $J$=~~8--~~7                   & ~~147& ~~690.552  & 17 USB &  13.4 (1.08)   &~~~~38  & ~~1.19  & g-fitted     & \\
                                    & $J$=14--13                     & ~~435&  1207.853  & 05 USB &  88.8 (0.99)   &~~~120  &         &              & \\
\\
PN 
       $v$=0                        & $N$=13--12                     & ~~205& ~~610.588  & 13 LSB & ~~5.4 (0.99)   &~~~~22  & ~~0.44  & g-fitted     & \\
                                    & $N$=16--15                     & ~~307& ~~751.311  & 11 LSB &  13.5 (1.00)   &~~~~39  & ~~0.64  & g-fitted     & \\                                                                                                                                                              
       $v$=1                        & $N$=12--11                     &  2078& ~~559.665  & 14 LSB & ~~6.6 (1.08)   &~$\lesssim$20&$\lesssim$0.16~ \\ 
                                    & $N$=25--24                     &  2631&  1164.391  & 06 USB &  59.5 (1.03)   &$\lesssim$179&$\lesssim$1.39~ \\
\\
\hline
\end{tabular}
\tablefoot{
$^\ddagger$ For the spectral resolution in \kms\ given in parentheses.}
\end{table*}                                                                                                                                                                      

In addition to these ground-vibrational lines, we detected some SO lines from
the first vibrationally excited state $v$=1, the \jkul{J}{N}{(J{\rm-1})}{(N-1)} triplets 
$J$=26--25 and $J$=27--26 at 1108 and 1150 GHz, and one component (the other two are blended 
with a very strong water line) of the $J$=23--22 triplet at 980.531\,GHz. The upper levels of these lines
lie between  2178 and 2386\,K above the ground. As in the case of SiO, when comparing the rotational diagram
of these $v$=1 transitions with those from the $v$=0, we see that the rotational temperature
is very similar in spite of the much higher excitation energy of the $v$=1 levels. In contrast,
when we compare the intensity of similar rotational lines in the two vibrationally excited states,
we derive vibrational excitation temperatures of about 2000\,K. This result points out that 
the excitation of the vibrationally excited states is mostly due to radiative pumping for SO as well.

The profiles of the well-detected lines are triangular or Gaussian-like, see Fig.\,\ref{fig-so}. 
The FWHM values obtained for the central component show a
clear decreasing trend with increasing excitation energy of the levels, ranging from 35\,\kms\ for
the $v$=0 $J$=13--12 lines to 20\,\kms\ for the $v$=0 \mbox{$J$=23--22} and all the $v$=1 detected lines
(though in these latter cases the S/N is poor and the results of the fittings are less 
accurate). 

\subsection{Sulphur dioxide (\sodos)}

\begin{figure}
	\resizebox{\hspec}{!}{\includegraphics{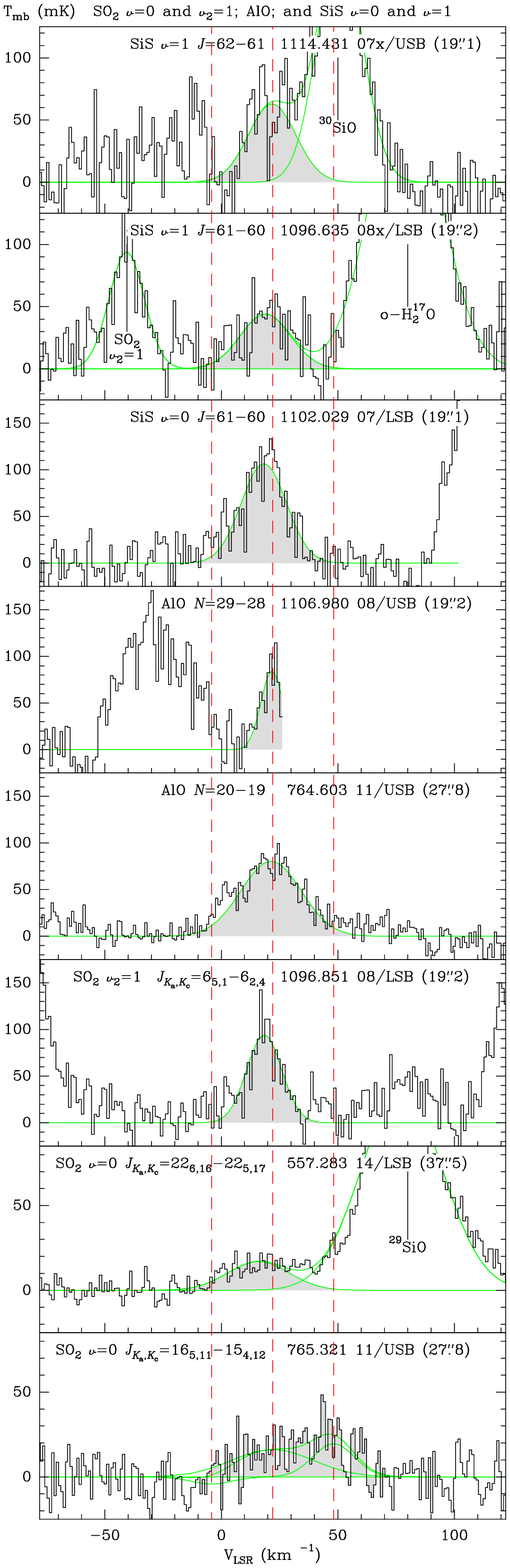}} 
	\caption{Same as Fig.\,\ref{fig-co} for the detected lines of AlO, and SiS, and some lines of \sodos.
	The assignation of the \nudu\ \jkk=\jkkul{6}{5}{1}{6}{2}{4} line of \sodos\ is tentative.}
	\label{fig-o-others}
\end{figure}

\begin{table*}
\caption{Spectral-line results for sulphur dioxide.}             
\label{tabso2}              
\begin{tabular}{rlccccccccc}    
\hline\hline 
Species and & Rotational    & $E_{\rm upp}/k$  
                                                                     & Rest freq.       & Setting \& &  ~r.m.s.$^\ddagger$   
                                                                                                                 & Peak & Area    & Comments \\
elec./vibr. state & quantum nums. &  (K)                                        &  (GHz)           & sideband&  (mK)        & (mK) & (K\,\kms)\\
\hline
\\
\sodos\    
   $v$=0   & \jkk=\jkkul{10}{6}{4~\,}{~~9}{5}{5}    & ~~139& ~~753.060  & 11 LSB &  13.6 (1.00)   &~~~~34 & ~~1.50  & \\
             & \jkk=\jkkul{16}{5}{11}{15}{4}{12} & ~~186& ~~765.321  & 11 USB &  13.5 (0.98)   &~~~~24 & ~~1.05  & \\
             & \jkk=\jkkul{~~8}{8}{0~\,}{~~8}{7}{1}     & ~~190& ~~763.982  & 11 USB &  13.5 (0.98)   & \multicolumn{2}{c}{(a)}      & \\ 
             & \jkk=\jkkul{~~9}{8}{2~\,}{~~9}{7}{3}     & ~~198& ~~763.992  & 11 USB &  13.5 (0.98)   & \multicolumn{2}{c}{(a)}      & \\
             & \jkk=\jkkul{10}{8}{2~\,}{10}{7}{3}   & ~~207& ~~763.999  & 11 USB &  13.5 (0.98)   & \multicolumn{2}{c}{(a)}      & \\
             & \jkk=\jkkul{16}{6}{10}{16}{5}{11} & ~~213& ~~560.614  & 14 LSB & ~~6.6 (1.08)   &~~~~11 & ~~0.79  & \\
             & \jkk=\jkkul{11}{8}{4~\,}{11}{7}{5}   & ~~217& ~~764.000  & 11 USB &  13.5 (0.98)   & \multicolumn{2}{c}{(a)}      & \\
             & \jkk=\jkkul{12}{8}{4~\,}{12}{7}{5}   & ~~229& ~~763.993  & 11 USB &  13.5 (0.98)   & \multicolumn{2}{c}{(a)}      & \\
             & \jkk=\jkkul{13}{8}{6~\,}{13}{7}{7}   & ~~240& ~~763.976  & 11 USB &  13.5 (0.98)   & \multicolumn{2}{c}{(a)}      & \\
             & \jkk=\jkkul{16}{7}{9~\,}{15}{6}{10}  & ~~245& ~~969.224  & 10 LSB &  19.8 (1.08)   &~~~~30 & ~~1.26  &  \\
             & \jkk=\jkkul{18}{6}{12}{18}{5}{13} & ~~246& ~~559.882  & 14 LSB & ~~6.6 (1.08)   &~~~~~~8 & ~~0.31 &  \\
             & \jkk=\jkkul{14}{8}{6~\,}{14}{7}{7}   & ~~253& ~~763.948  & 11 USB &  13.5 (0.98)   &  \multicolumn{2}{c}{(a)}     & \\
             & \jkk=\jkkul{17}{7}{11}{16}{6}{10} & ~~261& ~~988.302  & 09 LSB &  33.1 (1.06)   &~~~~23 & ~~045   & \\
             & \jkk=\jkkul{19}{6}{14}{19}{5}{15} & ~~263& ~~559.500  & 14 LSB & ~~6.6 (1.08)   &~~~~~~8 & ~~0.18 & \\
             & \jkk=\jkkul{15}{8}{8~\,}{15}{7}{9}   & ~~267& ~~763.906  & 11 USB &  13.5 (0.98)   &  \multicolumn{2}{c}{(a)}     & \\
             & \jkk=\jkkul{20}{6}{14}{20}{5}{15} & ~~281& ~~558.813  & 14 LSB & ~~6.6 (1.08)   &~~~~~~8 & ~~0.18 &  \\
             & \jkk=\jkkul{16}{8}{8~\,}{16}{7}{9}   & ~~282& ~~763.847  & 11 USB &  13.5 (0.98)   &  \multicolumn{2}{c}{(a)}     & \\
             & \jkk=\jkkul{17}{8}{10}{17}{7}{11} & ~~297& ~~763.768  & 11 USB &  13.5 (0.98)   &  \multicolumn{2}{c}{(a)}     & \\
             & \jkk=\jkkul{21}{6}{16}{21}{5}{17} & ~~301& ~~558.391  & 14 LSB & ~~6.6 (1.08)   &~~~~~~9 & ~~0.30 & \\
             & \jkk=\jkkul{20}{7}{13}{20}{6}{14} & ~~313& ~~661.668  & 12 USB & ~~7.8 (0.91)   &~~~~13 & ~~0.38  & \\
             & \jkk=\jkkul{18}{8}{10}{18}{7}{11} & ~~314& ~~763.668  & 11 USB &  13.5 (0.98)   &  \multicolumn{2}{c}{(a)}     & \\
             & \jkk=\jkkul{22}{6}{16}{22}{5}{17} & ~~321& ~~557.283  & 14 LSB & ~~6.6 (1.08)   &~~~~16 & ~~0.52  &  \\
             & \jkk=\jkkul{22}{7}{15}{22}{6}{16} & ~~352& ~~660.918  & 12 USB & ~~7.8 (0.91)   &~~~~21 & ~~0.30  & \\
             & \jkk=\jkkul{23}{7}{17}{22}{6}{16} & ~~374&  1102.115  & 07 LSB &  22.0 (1.09)   &~~~~25 & ~~1.05  & \\
             & \jkk=\jkkul{23}{7}{17}{23}{6}{18} & ~~374& ~~660.473  & 12 USB & ~~7.8 (0.91)   &~~~~12 & ~~0.15  & \\
             & \jkk=\jkkul{28}{4}{24}{27}{3}{25} & ~~416& ~~968.697  & 10 LSB &  19.8 (1.08)   &~~~~12 & ~~0.37  & \\
             & \jkk=\jkkul{25}{7}{19}{25}{6}{20} & ~~419& ~~659.338  & 12 USB & ~~7.8 (0.91)   &~~~~~~6 & ~~0.18 & \\
             & \jkk=\jkkul{26}{7}{19}{26}{6}{20} & ~~443& ~~658.542  & 12 USB & ~~7.8 (0.91)   &~~~~10 & ~~0.40  & \\
             & \jkk=\jkkul{32}{0}{32}{31}{1}{31} & ~~459& ~~571.553  & 14 USB & ~~6.6 (1.05)   &~~~~10 & ~~0.38  & \\
             & \jkk=\jkkul{32}{2}{30}{31}{3}{29} & ~~504& ~~571.533  & 14 USB & ~~6.6 (1.05)   &~~~~13 & ~~0.50  & \\
             & \jkk=\jkkul{34}{1}{33}{33}{2}{32} & ~~543& ~~622.827  & 13 USB & ~~5.4 (0.97)   &~~~~12 & ~~0.32  & \\
             & \jkk=\jkkul{37}{1}{37}{36}{0}{36} & ~~609& ~~659.421  & 12 USB & ~~7.8 (0.91)   &~~~~16 & ~~0.95  & \\
             & \jkk=\jkkul{36}{2}{34}{35}{3}{33} & ~~630& ~~661.511  & 12 USB & ~~7.8 (0.91)   &~~~~14 & ~~0.44  & \\ 
\nudu        & \jkk=\jkkul{~~6}{5}{1~\,}{~~6}{2}{4}     &~~843 &  1096.851  & 08 LSB &  23.1 (1.10)   &~~~~93  & ~~1.77  & maser ?\\
\\
\hline
\end{tabular}
\tablefoot{
$^\ddagger$ For the spectral resolution in \kms\ given in parentheses.
(a) Lines \jkk=$J_{8,J-8}$--$J_{7,J-7}$  for even values of $J$ between 8 and 18, and \jkk=$J_{8,J-7}$--$J_{7,J-6}$ 
for odd values of $J$ between 9 and 17, appear to be severely blended. They were fitted as a group, resulting
in a integrated area of 1.20\,K\,\kms\ and peak values between 10 and 20\,mK.}
\end{table*}    

\begin{table*}
\caption{Unassigned spectral features}             
\label{tabunknown}      
\begin{tabular}{cccccll}   
\hline\hline 
Rest freq. (GHz)       & Setting & ~r.m.s.$^\ddagger$       & Peak & Area     & \multicolumn{2}{c}{Possible assignations and comments}       \\
 LSB  / USB               & name  &  (mK)        & (mK) & (K\,\kms)&  \\ \hline   
559.675 (25) / 569.555 (25) & 14 & ~~6.6 (1.08) &   ~~22  & 0.40 & $^{29}$SiS $J$=32--31 & (569.528 GHz USB) \\
607.055 (25) / 622.774 (25) & 13 & ~~5.4 (0.99) &   ~~16  & 0.77 & $^{29}$SiS $J$=35--34 & (622.760 GHz USB); blended with SO$_2$ \\
751.664 (25) / 766.264 (25) & 11 &  13.6 (1.00) &   ~~60  & 1.40 & $^{33}$SO \jn=\jkul{18}{18}{17}{17} & (766.263 GHz USB)\\
\hline
\end{tabular}
\tablefoot{
$^\ddagger$ For the spectral resolution in \kms\ given in parentheses.}
\end{table*}

We tentatively detected 33 \sodos\ lines in our spectra,
see Table\,\ref{tabso2}. Almost all these lines are only slightly stronger that the detection limit,
therefore their parameters are rather uncertain. However, we verified that our
identifications are compatible with an unresolved, optically thin emission with a rotational excitation of about 250--300\,K.
Under these assumptions we checked that only the lines listed in the table are
visible in the spectra. Other lines may also be present, but they happen to overlap much
stronger lines of other species and therefore cannot be analysed. The detected transitions have 
high line-strengths and $E_{\rm upp}/k$ $\lesssim$\,630\,K.

In addition to these $v$=0 lines, we identified another spectral feature that we tentatively 
assign to \sotwo. For a systemic velocity of 22\,\kms\ and using a single-component Gaussian fitting, 
we derived rest frequencies of 1,096.860\,GHz (if from the LSB) and 1,109.215\,GHz (if from the USB) for the spectral 
feature
that appears to the left of the \nuz\ \jkkul{3}{1}{2}{3}{0}{3} line of ortho-\water\ in setting\,08 (Fig.\,\ref{fig-wbs-a}).
The spectral feature is clear (Fig.\,\ref{fig-o-others}), with a \tmb\ value of about 93\,mK. 
This line is also tentatively detected in the spectra of other
O-rich sources in HIFISTARS, such as \object{o\,\,Cet}, \object{W\,Hya}, and \object{R\,Dor}, and it is clearly absent from
the C-rich envelope around \object{IRC\,+10216}, and in the O-rich envelopes around \object{IK\,Tau}, 
\object{IRC\,+10011}, and \object{R\,Cas} (HIFISTARS team, priv. comm.). These observations also suggest that the
spectral feature arises from the LSB. We searched for lines in a 50\,MHz (13.5\,\kms) 
interval around the two possible frequencies in the Splatalogue catalogue 
(http://www.splatalogue.net/) and in the private line catalogues of some of
the members of the HIFISTARS team. According to this, the most likely candidate for our detection 
is the \nudu\ \jkk=\jkkul{6}{5}{1}{6}{2}{4}\ of \sotwo\ at 1096.851\,GHz. This line is only at
10\,MHz (2.7\,\kms) from the LSB value derived from the fit. Indeed, this is the only 
catalogued line in these frequency ranges from a species previously detected in \vycma. Nevertheless, the
identification of the line is tentative because the \jkkul{6}{5}{1}{6}{2}{4} transition has
a very low line-strength, it is a $\Delta K$=3 transition with $\log_{10}(A_{ul}) = -6.85$, 
but the detected spectral feature is relatively intense.  In addition,
it is a vibrationally excited line, its upper level being placed at more than
800\,K above the ground. If plotted in a rotational diagram for
\sotwo\ lines, we observe that the detected intensity is five orders of magnitude higher
than expected for {\em LTE} conditions. We can only explain this result if we assume that our line shows a strong maser
effect. The FWHM of the line is 14\,\kms, a value similar to that of other 
lines arising from the inner envelope, such as those of NaCl and other vibrationally excited lines 
of \sodos\ \citep{ziurys2007,tenenbaum2010a,tenenbaum2010b}.
The line is slightly blue-shifted w.r.t. the systemic velocity
(as is also the case for \object{o\,\,Cet}), which could be explained if the maser is due to an IR line overlap. 

\subsection{Other oxides: phosphorous monoxide and aluminium monoxide (PO and AlO)}\label{POAlO}

\begin{figure}
	\resizebox{\hspec}{!}{\includegraphics{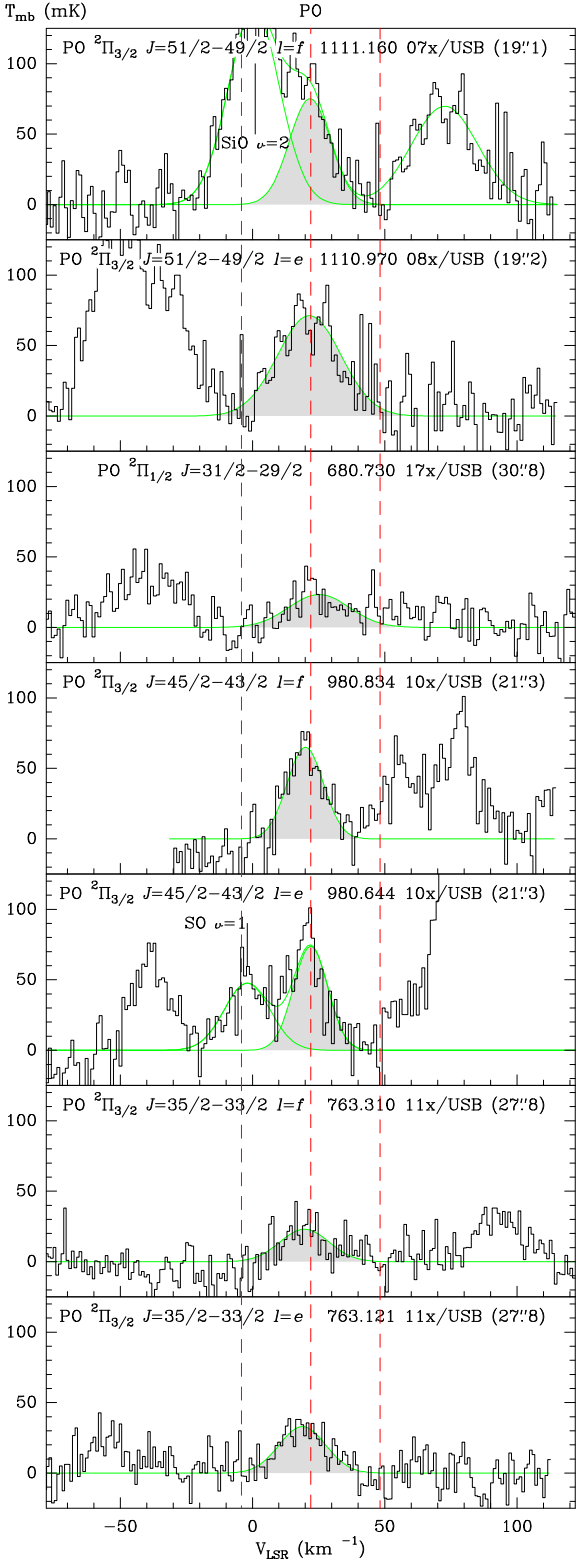}}  
	\caption{Same as Fig.\,\ref{fig-co} for the detected lines of PO.}
	\label{fig-po}
\end{figure}

We detected several lines of PO, both from the \twopihalf\ and \twopithalf\ electronic states.
The \twopithalf\ transitions always come in {\em paired features}, corresponding to the $l$=$e$ and $l$=$f$ $\Lambda$-doubling
components, while the \twopihalf\ always show {\em single features} because we cannot separate  
the $\Lambda$-doubling splitting in this case\footnote{We only consider the strongest hyperfine components, for 
\mbox{$J$=$j$--($j$--1)} the 
\mbox{$F$=($j$+1/2)--($j$--1/2)} and \mbox{$F$=($j$--1/2)--($j$--3/2)} components, 
which are not resolved in \vycma\ for both \twopihalf\ and \twopithalf\ electronic states.}. We clearly detected the 
\twopithalf\  \mbox{$J$=35/2--33/2}, \mbox{45/2--43/2}, and \mbox{51/2--49/2} pairs, with upper state energies at 338, 553, and 707\,K above the ground, 
respectively. Another two \twopithalf\ transitions, the \mbox{$J$=79/2--77/2} and 83/2--81/2  with $E_{\rm upp}/k$ of 1700 and 2100\,K, 
were also observed, but since they are blended with other stronger lines, we cannot conclusively determine their presence.
For the \twopihalf\ transitions, we only detected the $J$=31/2--29/2 at 589\,K above the ground.
The $J$=53/2--51/2 and 55/2--53/2 lines, at $E_{\rm upp}/k \sim$\,1100\,K, are also within the band, but again they are blended
with much stronger emission from other species. In all cases the lines only show the central main
component, see Fig.\,\ref{fig-po}. The rotational diagram suggests an excitation temperature of 400\,K (Fig.\,\ref{rotation-others}).

We also detected two lines of AlO: the \mbox{$N$=20--19} and \mbox{29--28} 
transitions with \Euppk\ of 386 and 798\,K. The \mbox{$N$=45--44} and \mbox{49--48} transitions were also observed,
but we did not detect them, 
the upper limits not being significant. 
The \mbox{$N$=29--28} line is detected at
the edge of the band in the setting\,08 observation, and should be considered as tentative
(see Fig.\,\ref{fig-o-others}). Because of this, the estimate of the excitation temperature
from the rotational diagram is very uncertain; we derive a value of 160\,K, a rather low
figure compared with the 230\,K found by \citet{tenenbaum2009}, but still compatible in view of
the uncertainties. The $N$=20--19 line is clearly detected, 
the profile just consisting in the central component. 

\subsection{Silicon sulphide (SiS)}

We detected only one $v$=0 rotational line of SiS, the \mbox{$J$=61--60} at 1102.029 GHz. 
The \mbox{$J$=104--103} at 1861.245\,GHz is observed in setting\,01, overlapping the 
vibrationally excited line of water at 1868.783\,GHz, but it is very unlikely that we could have detected
the SiS line since it has an upper state energy of 4723\,K above the ground. No other $v$=0 lines of SiS 
were observed, but several other rotational transitions of vibrationally excited 
states were covered by our survey. We did not detect any lines from states
$v$=2 and higher, but we tentatively detected two lines from the $v$=1 levels: the
$J$=\mbox{61--60} at 1096.635 GHz, and the $J$=\mbox{62--61} at 1114.431\,GHz 
\citep[this line has been also detected in OH\,26.5+0.6, see][]{justtanont2012}. In addition, we obtained
significant upper limits for the $v$=1 $J$=55--54 and $J$=64--63
transitions of SiS. 
The values obtained for the $v$=1 $J$=62--61 transition may be affected by the blending of this line
with the $v$=0 $J$=26--25 line of \trsio\ entering from the other side-band of the receiver,
but they are very similar to the adjacent $J$=61--60 line, which presents a clear detection.
These two lines show just about half the strength of the $v$=0 $J$=61--60, even though they
have an excitation energy of 2700\,K, that is, 1600\,K above the $v$=0 line. This would suggest a relatively 
high excitation temperature between the $v$=0 and $v$=1 ladders 
 due to the IR pumping of the $v$=1 levels. However, the non-detection of the 
$v$=1 $J$=55--54 and $J$=64--63 transitions suggests that the strength of the two detected $v$=1 lines 
may also be enhanced by an IR line overlap. 

\begin{figure}
	\resizebox{\hspec}{!}{\includegraphics{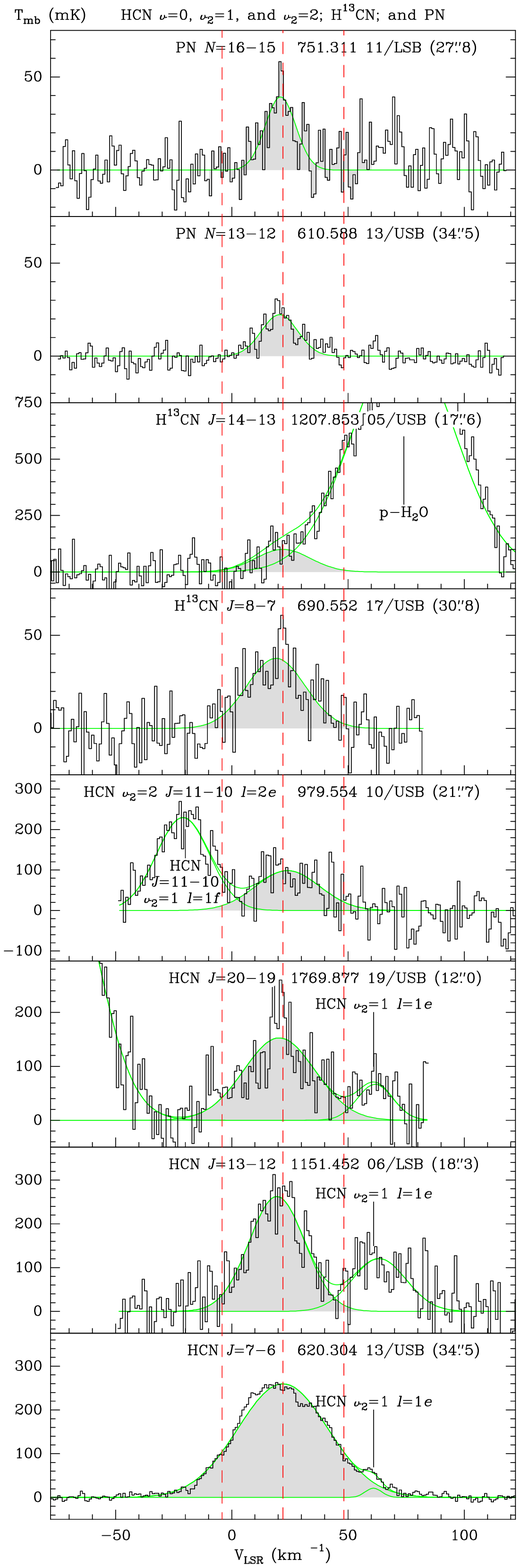}} 
	\caption{Same as Fig.\,\ref{fig-co} for the detected lines of HCN, \HtreceCN, and  PN.}
	\label{fig-n-others}
\end{figure}

\subsection{Hydrogen cyanide (HCN)}

\begin{table*}
\caption{Triple-Gaussian fitting results for CO, \water, \ammonia, SiO, SO, and \sodos.}             
\label{tabfits1}      
\begin{tabular}{rlccccccccc}   
\hline\hline 
   &           &      & \multicolumn{2}{c}{Central component} & \multicolumn{3}{c}{HEVW components} & \\ 
Species and &  Rotational      & $E_{\rm upp}^\dagger$  
                              & Peak       & FWHM                      &  Blue peak  & FWHM   & Red Peak      & \\
elec./vibr. state &  quantum nums. &    (K) & \tmb\ (mK) & (\kms)                    & \tmb\ (mK)  & (\kms) & \tmb\ (mK)    & \\ \hline
\\
\doceCO\ $v$=0   
          & $J$=~~6--~~5                   & ~~116& 1091       & 38    & ~~~~292&  19    & 366           & \\
          & $J$=10--~~9                    & ~~304& 1499       & 37    & ~~~~366&  22    & 452           & \\
          & \jdsq                          & ~~752& 1296       & 31    & ~~~~205&  17    & 264           & \\
\\
\treceCO\  $v$=0 
          & $J$=~~6--~~5                   & ~~111& ~~~~82     & 49    & ~~~~~~48&  12    &~~38           & \\
          & $J$=~~9--~~8                   & ~~238& ~~131      & 47    & \\
          & $J$=10--~~9                    & ~~291& ~~173      & 39    & \\
          & \jdsq                          & ~~719& ~~257      & 19    & \\              
\\
ortho-\water\ $v$=0  
          & \jkk=\jkkul{1}{1}{0}{1}{0}{1}  &~~~~27&  1143      & 43    & ~~$-$253    &  17    &~~92           & \\
          & \jkk=\jkkul{3}{0}{3}{2}{1}{2}  & ~~163&  4664      & 43    & $-$1514     &  17    & 538           & \\
          & \jkk=\jkkul{3}{1}{2}{3}{0}{3}  & ~~215&  1809      & 45    & ~~$-$268      &  21    & 232           & \\
          & \jkk=\jkkul{3}{1}{2}{2}{2}{1}  & ~~215&  2919      & 46    & ~~$-$425      &  13    & 310           & \\
          & \jkk=\jkkul{3}{2}{1}{3}{1}{2}  & ~~271&  2009      & 41    & ~~~~$-$25       &  25    & 364           & \\
          & \jkk=\jkkul{5}{3}{2}{5}{2}{3}  & ~~698&  1802      & 31    & \\
\nudu     & \jkk=\jkkul{2}{1}{2}{1}{0}{1}  & 2379 & 1007       & 32    & \\
          & \jkk=\jkkul{8}{2}{7}{8}{3}{4}  & 3556 & ~~~~46     & 16    & \\                                                                                                                                                              
\\
para-\water\ $v$=0 
          & \jkk=\jkkul{1}{1}{1}{0}{0}{0}  &~~~~53&  3140      & 43    & ~~$-$804      &  18    & 373           & \\
          & \jkk=\jkkul{2}{0}{2}{1}{1}{1}  & ~~101&  3317      & 40    & ~~$-$469      &  12    & 287           & \\
          & \jkk=\jkkul{2}{1}{1}{2}{0}{2}  & ~~137&  1570      & 45    & ~~$-$101      &  23    & 267           & \\
          & \jkk=\jkkul{4}{2}{2}{4}{1}{3}  & ~~454&  1183      & 50    &             &  16    & 167           & \\
          & \jkk=\jkkul{6}{3}{3}{6}{2}{4}  & ~~952&  1011      & 31    & \\
\nudu     & \jkk=\jkkul{1}{1}{1}{0}{0}{0}  & 2352 & ~~244      & 28    & \\
\nudd     & \jkk=\jkkul{2}{1}{1}{2}{0}{2}  & 4684 & ~~114      & 27    & \\
\nutu     & \jkk=\jkkul{6}{3}{3}{6}{2}{4}  & 6342 & ~~169      & 27    & \\
\\
ortho-\waterdo\ $v$=0 
          & \jkk=\jkkul{3}{1}{2}{3}{0}{3}  &~~215 &   209      & 34    & \\
\\                                                                                                                                                               
para-\waterdo\ $v$=0 
          & \jkk=\jkkul{1}{1}{1}{0}{0}{0}  &~~~~53&   205      & 35    & ~~~~$-$62       & 19     & ~~47            & \\
\\                                                                                                                                                               
ortho-\waterds\ $v$=0 
          & \jkk=\jkkul{3}{0}{3}{2}{1}{2}  &~~162 &   350      & 26    & \\
          & \jkk=\jkkul{3}{1}{2}{3}{0}{3}  &~~215 &   232      & 35    & \\
\\                                                                                                                                                               
para-\waterds\ $v$=0 
          & \jkk=\jkkul{1}{1}{1}{0}{0}{0}  &~~~~53&   126      & 38    & ~~~~$-$49       & 11     & ~~$-$9          & \\
          & \jkk=\jkkul{2}{0}{2}{1}{1}{1}  &~~101 & ~~~~72     & 35    & \\
\\                                                                                                                                                                                                                                                                                                                                          
ortho-\ammonia\  $v$=0 
          & \jk=\nhtuz--\nhtzz             &~~~~27& ~~235      & 41    & ~~~~~~89         & 19     & 183           & \\
          & \jk=\nhttz--\nhtdz             & ~~170&  1441      & 45    & \\
\\
para-\ammonia\  $v$=0 
          & \jk=\nhttd--\nhtdd             & ~~127& ~~450      & 44    & \\
\\                                                                                                                                                               
\hydroxyl\ \twopihalf\ $v$=0  
          & \jtmdm                         &~~270 &  1695      & 40    & ~~~~122           & 19     & 641           & \\
\\
\vosio\ $v$=0 & $J$=14--13                     &~~219 & ~~369      & 32    & ~~~~~~41          & 25     & ~~65            & \\
          & $J$=16--15                     &~~283 & ~~378      & 28    & ~~~~~~43          & 23     & ~~90            & \\
\\                                                                                                                                                               
\vnsio\ $v$=0    
          & $J$=13--12                     &~~187 & ~~118      & 41    &  ~~~~~~12          &  25     &  ~~20            &  \\
          & $J$=26--25                     &~~722 & ~~203      & 32    & \\
\\                                                                                                                                                               
\trsio\ $v$=0   
          & $J$=26--25                     &~~713 & ~~193      & 23    & \\
          & $J$=42--41                     & 1832 & ~~153      & 22    & \\
\\
SO $v$=0  & \jn=\jkul{13}{14}{12}{13}      &~~193 & ~~~~70     & 32    & ~~~~~~~~2          &  24     &  ~~26            & \\
          & \jn=\jkul{13}{12}{12}{11}      &~~194 & ~~~~60     & 31    & ~~~~~~~~0          &  24     &  ~~20            & \\
          & \jn=\jkul{13}{13}{12}{12}      &~~201 & ~~~~64     & 29    & ~~~~~~10           &  21     &  ~~30            & \\
          & \jn=\jkul{15}{16}{14}{15}      &~~253 & ~~~~75     & 35    & ~~~~~~~~5           &  13     &  ~~10            & \\
\\
\sodos\  $v$=0  
       & \jkk=\jkkul{16}{5}{11}{15}{4}{12} &~~186 & ~~~~16     & 29    &  ~~~~~~$-$4        &  14     &  ~~19            &  \\ 
\\
\hline
\end{tabular}
\tablefoot{
$^\dagger$ Since the ortho and para spin isomer variants of both water and ammonia behave as different chemical species, for all ortho 
transition \Euppk\ values are relative to the corresponding lowest ortho energy level, while for the 
para transitions they are given w.r.t. the corresponding lowest para energy level.} 
\\         
\end{table*}  

We detected all the observed
rotational transitions of HCN from the ground vibrational state:  
the $J$=7--6, 13--12, and 20--19 lines, with \Euppk\
of 119, 387, and 893\,K. 
We also detected, 
although tentatively in some cases, all the HCN rotational lines within our bands from the 
vibrationally excited states 
\nudu\ and \nudd: the $J$=7--6, 13--12, and 20--19 $l$=1$e$ transitions of 
the \nudu\ state, and the \nudu\ $J$=11--10 $l$=1$f$ and \nudd\ $J$=11--10 $l$=2$e$ lines. 
Two other lines from the \nuuu\ and \nutu\ states 
were also observed, but were not detected.

The profiles of all these lines show only the central component.
Including the data for vibrationally excited transitions, we found a noisy trend when plotting the
FWHMs of the lines against the excitation energy of the upper levels. Lines with relatively low excitation
energy, for instance, the $v$=0 $J$=7--6, show FWHM values higher than 40\,\kms, while 
values lower than 20\,\kms\ are found for transitions with $\ge E_{\rm upp}/k$ of 1500\,K. This result 
suggests that the vibrationally excited lines of HCN arise from the inner regions of the 
envelope where the final expansion velocity has not been attained yet. 

The rotational diagram for HCN shows excitation temperatures between 90 and 250\,K for
the $v$=0 lines, and of about 200\,K for the $v_2$=1 lines (Fig.\,\ref{rotation-others}). If we compare similar rotational
transitions in different vibrational states, we obtain excitation temperatures of 
210--350\,K between the $v$=0 and $v_2$=1 ladders (and of 1500\,K between the
$v_2$=1 and $v_2$=2 ladders, but this a very tentative result). This would suggest, although less clearly than
in the cases of \water, SiO, and SO, that the excitation of the $v_2$=1 and 2 
levels in HCN is also dominated by the IR radiation pumping.

We also systematically searched for lines of \HtreceCN\ in our spectra, both from the ground and the \nudu\
states.
We detected the \nuz\ $J$=8--7 and (tentatively) the $J$=14--13 lines, 
with \Euppk\ of 145 and 435, respectively. No vibrationally excited lines were 
detected in our data, which include the frequencies of the \nudu\ $J$=7--6 and
$J$=8--7 transitions.
Because of the severe blending of the \nuz\ $J$=14--13 line (it is
at the edge of the strong \jkkul{4}{2}{2}{4}{1}{3} line of para-\water), the results from the rotational diagram of 
\HtreceCN\ can only be considered as tentative. We derive an excitation temperature of about 90\,K,
very similar to the value for HCN between similar transitions, suggesting that the lines 
from both species are optically thin.
 
\subsection{Phosphorous nitride (PN)}\label{sec:pn}

In addition to \ammonia\ and HCN, the other N-bearing molecule identified in \vycma\ in our
HIFI spectra is PN. We detected the only two (ground-vibrational state) rotational 
lines covered by our observations, the $N$=13--12 and 16--15 multiplets (six unresolved components within 4\,MHz), with upper 
level energies of 205 and 307\,K above the ground. The lines are relatively narrow, with FWHPs of less than 
20\,\kms, suggesting that the emission originates in the inner shells of the envelope. 
The excitation temperature derived from the rotational diagram is about 220\,K.
Two $v$=1 multiplets, the $N$=12--11 and 25--24 at 559.7 and
1164.4\,GHz, were also observed but were not detected, the limits  are irrelevant.

\subsection{Unassigned spectral features}\label{sec:non}

There are three spectral features in our spectra that we were unable to unambiguously
assign to any molecular line. They are listed in Table\,\ref{tabunknown} and are marked
'U' in green in Fig.\,\ref{fig-wbs-a}. For the two in settings\,13 and 14,
$^{29}$SiS is a good candidate. For the other U-line in setting\,11,
the most likely candidate is $^{33}$SO (\jn=\jkul{18}{18}{17}{17} at 766.263\,GHz).
In other HIFISTARS observations, the U-features in settings\,13 and 14 are also detected in C-rich envelopes, whereas
that in setting\,11 seems to be present in other O-rich sources, which would support our conjectures.

\section{Full Herschel/HIFI WBS spectra}\label{sec:full-wbs-spectra}

\begin{figure*}
	\resizebox{16.04cm}{!}{\includegraphics{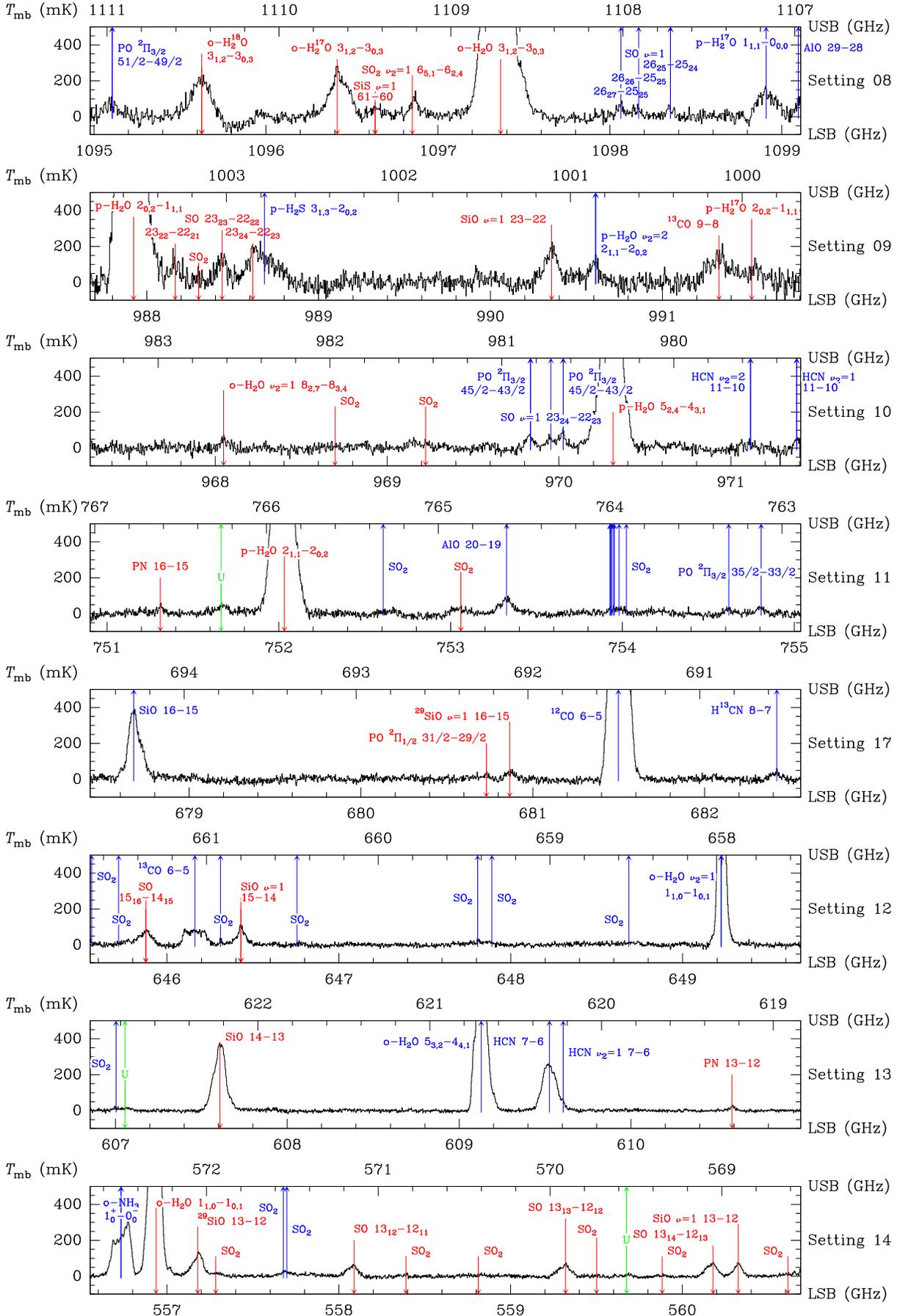}}
	\caption{Full WBS bandwidth HIFI results for \vycma\ for LO frequencies of 
	1102.9 GHz and lower. Quantum numbers for \sodos\ transitions are not given for clarity.}
	\label{fig-wbs-a}
\end{figure*}

\begin{figure*}
	\resizebox{16.04cm}{!}{\includegraphics{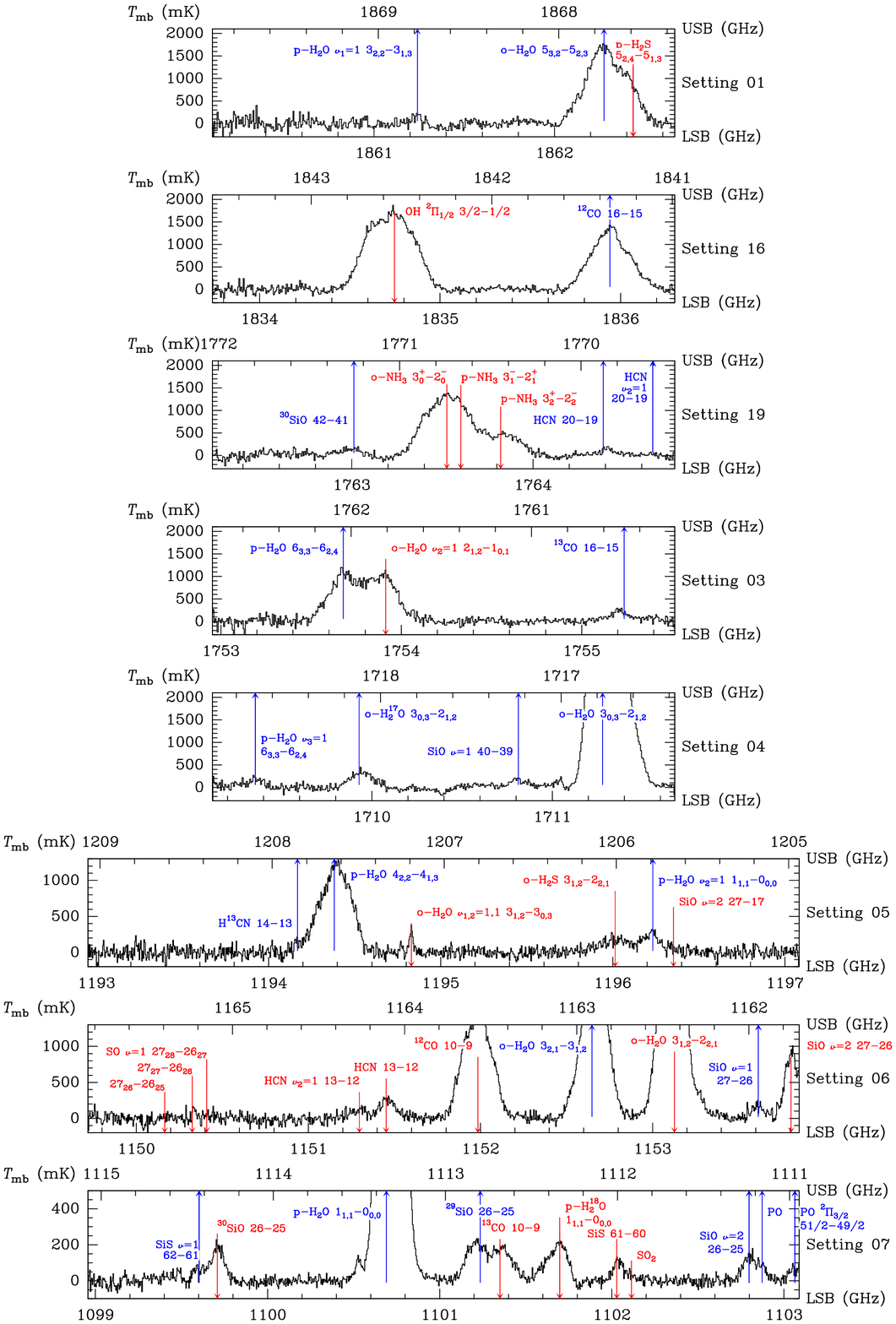}} 
	\caption{Full WBS bandwidth HIFI results for \vycma\ for LO frequencies of  
	1106.9 GHz and higher. Quantum numbers for \sodos\ transitions are not given for clarity.}
	\label{fig-wbs-b}
\end{figure*}

Here we present the full-bandwidth results of our Herschel/HIFI WBS observations of 
\vycma.  In Fig.\,\ref{fig-wbs-a} we show the results for LO frequencies of 
1102.9 GHz and lower: these are, from bottom to top, settings 14, 13, 12, 17, 11, 10, 09,
and 08. In Fig.\,\ref{fig-wbs-b} we show the results for LO frequencies of 
1106.9 GHz and higher: these are, from bottom to top, settings 07, 06, 05, 04, 03, 19, 16, and 01.
The \tmb\ observed with HIFI/WBS is plotted vs. both LSB (lower x-axis) and USB (upper x-axis) 
rest-frequency scales (in GHz), assuming an \LSR\ systemic velocity of 22\,\kms\ for the source. 
The location of the detected lines from the LSB/USB is indicated by the red and blue arrows and labels. 
The point of the arrow marks the appropriate axis in each case. Unassigned features are
labelled  `U' in green, and arrows point to the two possible rest frequencies. 
We note that setting 16 was observed twice using a slightly different {\em LSR} velocity 
for the source; since we cannot average these two spectra simultaneously aligning the two (side-band) 
frequency scales, in Fig\,\ref{fig-wbs-b} we only show the result for OBSID 1342196570.

\end{appendix}

\end{document}